\documentclass[twocolumn]{aastex63} 

\usepackage[colorinlistoftodos]{todonotes}
\usepackage[flushleft]{threeparttable}
\usepackage{graphicx}

\let\Oldtodo\todo
\renewcommand{\todo}[1]{\Oldtodo[inline]{#1}}

\hypersetup{linkcolor=cyan,citecolor=cyan,filecolor=cyan,urlcolor=cyan}

\usepackage{amsmath}
\usepackage{float}

\shorttitle{Coupled Interiors \& Tidal Dynamics}
\shortauthors{Hallatt \& Millholland}

\graphicspath{{./}{figures/}}

\begin{document}

\title{Coupled Planetary Interior and Tidal Evolution}

\correspondingauthor{Tim Hallatt}
\email{thallatt@mit.edu}

\author[0000-0003-4992-8427]{Tim Hallatt}
\affil{MIT Kavli Institute for Astrophysics and Space Research, Massachusetts Institute of Technology, Cambridge, MA 02139, USA}

\author[0000-0003-3130-2282]{Sarah Millholland}
\affil{MIT Kavli Institute for Astrophysics and Space Research, Massachusetts Institute of Technology, Cambridge, MA 02139, USA}
\affil{Department of Physics, Massachusetts Institute of Technology, Cambridge, MA 02139, USA}

\begin{abstract}
We present a new planetary structure/thermal evolution model, designed for use in problems that couple orbital dynamics with planetary structure. We first benchmark our structural/thermal evolution calculations against the \texttt{MESA} stellar evolution code, finding excellent agreement across a wide range of planet mass, equilibrium temperature, entropy, and extra heating deposited at various depths in the planet. We then apply our method to study the tidal migration histories of Neptunes in the recently identified ``ridge" (periods ${\sim}3{-}6$ days), a feature that has been suggested to be populated via high eccentricity migration (HEM) of more distant Neptunes. We find that it is difficult to form a circularized Neptune in the ridge without instigating runaway tidal inflation and likely atmospheric destruction; low eccentricity Neptunes in the ridge can only be emplaced by HEM if they are metal-rich and exhibit finely-tuned tidal quality factors. If follow-up observations confirm that low eccentricity Neptunes in the ridge did arrive via HEM and are not strongly enriched in metals, our calculations indicate that their tidal heating mechanism must operate in the upper reaches of the planet to avoid runaway inflation. Gravity modes excited in upper radiative layers are a possible candidate mechanism, while friction in the core or turbulent dissipation in convective zones could be ruled out.
\end{abstract}

\section{Introduction} \label{sec:intro}

The tidal evolution of planets is intimately connected to their interior structure: planetary structure dictates how tides operate in the planet, which in turn alter the structure \citep[e.g.][]{ogi14}. Despite this tidal/structural coupling, a common simplification adopted in tidal evolution studies is to reduce planets to point particles and thereby treat tidal evolution in isolation. While neglecting tidal/structural feedback has been fruitful, there is a growing consensus that endowing planets with structure can have far-reaching implications for planetary tidal evolution histories. The goal of this paper is to address this shortcoming of current tidal evolution calculations by developing an easy to use, fast, and accurate method to couple planetary dynamics (including tides) to planetary structure.

The chief mechanism by which tidal/structural feedback alters planets' evolution histories is via tidal heating: the conversion of orbital energy to heat in the planetary interior can inflate planetary radii, which in turn alters the rate of tidal processes \citep[e.g. the tidal luminosity, and eccentricity and semi-major axis damping rates ${\propto}R^{5}_{\rm p}$; e.g.][]{lecchabar10}. Giant planets on eccentric orbits can indeed circularize to become a hot Jupiter (i.e. undergo high eccentricity migration, HEM) up to an order of magnitude more rapidly if tidal/structural coupling is accounted for, due to early radius inflation \citep[][]{ibgbur09,milforjac09,rozglaper22}. Tidal radius inflation  may also change the outcome of HEM by quenching the eccentricity and inclination oscillations that can trigger migration \citep[i.e. von Zeipel-Lidov-Kozai (ZLK) oscillations;][]{fabtre07,itooht19} due to increasing rates of tidal precession \citep[which are also ${\propto}R^{5}_{\rm p}$;][]{luanli25}. Such dynamical/structural coupling could potentially reconcile the hot Jupiter period distribution predicted by HEM theory with the data \citep[][]{pet15a}. Planets' structural response to tides may also explain the anomalously puffy structure and mass loss signatures of the hot Neptune WASP 107 b \citep[][]{yudai24} and other Neptunes on polar orbits \citep[][]{setmil25}. Structural feedback may also account for lower mass sub-Neptunes just wide of resonance that are observed to be inflated \citep[][]{mil19,xudai25}. Tidal/structural feedback is therefore likely to operate across a broad range of planets and orbital architectures.

A common limitation of many (but not all) of the aforementioned studies is the use of simplified calculations of the planetary structural response to tides. While coupled structure and tidal calculations have been achieved for giant planets \citep[see e.g.][who each use different interior structure codes for this purpose]{ibgbur09,milforjac09,lecchabar10,glarozper22}, self-consistent tidal evolutionary calculations have not been undertaken for lower mass Neptune/sub-Saturn planets. Instead, the common approach has been to parameterize the degree of radius inflation, e.g. by using the hot Jupiter radius versus irradiation flux relation of \cite{thoforlop21} \citep[including a lag time to emulate the effect of thermal inertia on the planetary radius;][]{yudai24}, or the static calculations of radius versus internal luminosity performed by \cite{milpetbat20}. Another simplified approach has been to track radius inflation solely due to the evolving stellar irradiation during dynamical evolution, without including tidal heating \citep[][]{attbouegg21}.

These simplified approaches to coupling tides and planet structure fall short of more realistic calculations in three main respects. First, methods that parameterize radius inflation do not account for the thermal inertia of planets undergoing heating and cooling. This lag time between heating (cooling) and radius inflation (shrinkage) matters because of the extreme sensitivity of tidal processes to the instantaneous planetary radius \citep[][]{lecchabar10}. Second, planetary structural changes are also accompanied by changes in the Love number $k_{2,\rm p}$, which helps dictate the efficiency of tidal dissipation \citep[in standard, equilibrium tide theory; e.g.][]{hut81}. Despite the drastic changes in planet structure found to result from tidal heating \citep[e.g.][]{mil19}, concomitantly drastic changes in planetary Love number are rarely accounted for. Lastly, the degree to which a planet puffs up or shrinks depends on a host of structural parameters, e.g. equilibrium temperature, metallicity, entropy, and core mass. Despite the fact that each of these parameters helps govern the planetary response to tidal heating, they are seldom included in simplified tidal/structure calculations. 

This paper serves two purposes. Our first goal is to build and validate a new planet structure/thermal evolution model that is capable of self-consistently coupling planet structural and dynamical evolution. The approach is designed to be fast, accurate, and easy to implement into dynamical integrations. Our goals differ from some recently developed planet structure codes \citep[][]{knihel24,sursutej24} in that our method treats fewer physical effects (e.g. chemical diffusion, mixing, semiconvection) to streamline implementation into dynamical calculations while preserving first-order accurate planet structure physics. Our methodology covers a wide range of planet masses, metallicities, and extra heating prescriptions.

The second aim of this work is to showcase our method by studying the tidal migration of planets in the newly identified Neptunian ``ridge" (periods ${\sim}3{-}6$ days), which recent work suggests may have arrived via HEM \citep[][]{casboulil24,doyarmacu25}. We will show that coupling close-in Neptunes' internal structures to their tidal migration histories could yield a novel constraint on the tidal dissipation mechanism at work in their interiors. A second paper in this series (hereafter referred to as Paper II) further showcases our method, in which we explore the fate of hot Jupiters and the formation of planets deep in the sub-Jovian desert (periods ${\lesssim}3$ days). We have made our planet structure models available to the community upon publication of this series of papers.\footnote{\href{Available here}{https://github.com/thallatt/PSAND.git}\label{footnote:git}}

This paper is organized as follows. In Section \ref{sec:methods} we outline how we construct static and evolving planet structure models with and without extra heating. Section \ref{sec:methods} also details the microphysics employed in the planet structure calculations. We then benchmark our calculations against full fledged planetary evolution simulations with \texttt{MESA} in Section \ref{sec:mesa}. In Section \ref{sec:ridge} our method is applied to study the tidal histories of Neptunes in the ``ridge". We close this paper in Section \ref{sec:conclusion} by briefly summarizing our benchmarking results and the implications of our structure/dynamics calculations for the migration of close-in Neptunes.

\section{Methods}\label{sec:methods}

\subsection{Static Planets}\label{subsec:static_planets}

To follow the simultaneous internal and dynamical evolution of planets, we create a look-up table of planet internal structures spanning a range of masses, external irradiation fluxes, and internal entropies to be used on the fly during the dynamical integrations. For a given planet interior model, we solve the following stellar structure equations \citep[][]{cha39,coxgiu68}:

\begin{align}\label{equation:stellar_structure}
\begin{split}
    \frac{dm}{dr}&=4\pi r^{2}\rho(P,T)\\
    \frac{dP}{dr}&=-\frac{G m(<r)}{r^{2}}\rho(P,T)\\
    \frac{dT}{dr}&=\frac{dP}{dr}\frac{T}{P}\nabla(P,T),
\end{split}
\end{align}

\noindent with $r$ the radius, $m$ the mass enclosed within $r$, $\rho$ the density, $P$ the pressure, $T$ the temperature, and $\nabla{=}d\log T/d\log P$ the logarithmic temperature gradient with respect to pressure. We use the Schwarzschild criterion \citep{sch58} to determine whether convection or radiative diffusion set the local temperature gradient. The Schwarzschild criterion requires setting $\nabla{=}\mathrm{min}(\nabla_{\rm ad},\nabla_{\rm rad})$, with the adiabatic and radiative temperature gradients given by, respectively,

\begin{align}\label{equation:nablas}
\begin{split}
    \nabla_{\rm ad}&=-\frac{\partial \log S}{\partial \log P}\bigg|_{T}\bigg(\frac{\partial \log S}{\partial \log T}\bigg|_{P}\bigg)^{-1}\\
    \nabla_{\rm rad}&=\frac{3\kappa(P,T) P}{64 \pi G m \sigma_{\rm SB}T^{4}}L,
\end{split}
\end{align}

\noindent where $S$ is the specific entropy, $\kappa$ is the opacity, $G$ is the gravitational constant, $\sigma_{\rm SB}$ is the Stefan-Boltzmann constant, and $L$ is the luminosity. We assume throughout our calculations that the luminosity is spatially constant and flows outward ($L{>}0$). This assumption of spatially constant $L$ is violated if the planet does not maintain thermal equilibrium, which can occur if the thermal relaxation time in the radiative zone $t_{\rm rad}{\sim}c_{\rm p}T\rho H/F{\sim}P c_{\rm P} T/g F$ (with $c_{\rm P}$ the specific heat capacity, $\rho$ the density, $H{=}c^{2}_{\rm s}/g$ the scale height, $c_{\rm s}$ the sound speed, $g$ the gravitational acceleration, and $F$ the outgoing flux) is longer than the global cooling time of the planet $t_{\rm cool}{\sim}S/|dS/dt|$ (see equation \ref{equation:cooling}). If $t_{\rm rad} {\gtrsim} t_{\rm cool}$, energy cannot flow outward at constant rate through the radiative zone as the planet evolves. We have explicitly confirmed a posteriori that $t_{\rm rad} {\ll} t_{\rm cool}$ throughout our evolution calculations, justifying this constant $L$ assumption. 

Solving equations \ref{equation:stellar_structure} requires an equation of state to supply $\rho(P,T)$, $S(P,T)$, and $\nabla_{\rm ad}(P,T)$, a model for the opacity $\kappa(P,T)$, as well as boundary conditions for the top and bottom of the atmosphere. We detail our equation of state and opacities in Sections \ref{subsec:eos} and \ref{subsec:kappa}, respectively. We close the present section by outlining how we build a planet structure model of a given mass, luminosity, equilibrium temperature, and core mass that satisfies our chosen boundary conditions. We then discuss in Section \ref{subsection:thermal} how we compute cooling evolution from these models.

\begin{deluxetable*}{CCCCCCc}\label{tab:structure_tables}
\tablecaption{Parameters varied in our planet structure calculations. 
\label{tab:parameters}}
\tablecolumns{5}
\tablewidth{0pt}
\tablehead{
\colhead{Parameter} &
\colhead{Definition} &
\colhead{Value} & 
\colhead{Units} & 
}
\startdata
$m_{\rm p}$ & \rm planet \ mass & [10.1,3{\times}10^{2}] & M_{\oplus} \\
$m_{\rm c}$ & \rm \ planet \ core \ mass & $\{10,20\}$ & M_{\oplus} \\
$X,Z$ & \rm atmospheric \ mass \ fractions & $\{0.74,0.02\}$, \ $\{0.65,0.1\}$, \ $\{0.36,0.5\}$ & \nodata \\
$S$ & \rm entropy & [6${-}$11] \ (\mathrm{for} \ $Z{=}0.02,0.1$), \ [4${-}$8] \ ($Z{=}0.5$) & k_{\rm B}/m_{\rm H} \\
$T_{\rm eq}$ & \rm equilibrium \ temperature & $\{2884,2039,1442,912,288\}$ & K \\
\\
\enddata

\end{deluxetable*}

We define the planet photospheric radius $R_{\rm p}$ to be where the atmosphere becomes optically thick to its own thermal radiation. This occurs at a pressure $P_{\rm p}{=}2g/3\kappa(P_{\rm p},T_{\rm p})$, where $T_{\rm p}$ is the photospheric temperature. Given a choice of $m_{\rm p}$ and $L$, we guess initial values of $T_{\rm p}$ and $R_{\rm p}$. Equations \ref{equation:stellar_structure} are then integrated inward to the bottom boundary, which we take to be the surface of a core of mass $m_{\rm c}$ and radius $r_{\rm c}{=}(m_{\rm c}/M_{\oplus})^{0.25}$ \citep[matching the radii of Earth-like composition planets in our chosen regime of core mass within ${<}2\%$ of Figure 2 from][]{formarbar07}. For a given integration of equations \ref{equation:stellar_structure}, we use scipy's \texttt{brentq} root finder to locate $P_{\rm p}$ given the run of $T(r)$ and $P(r)$. The top boundary condition we enforce is that the photospheric temperature $T_{\rm p}$ matches that of the Eddington two-stream $T(\tau{=}2/3)$ solution of equation 29 in \cite{gui10}:

\begin{align}\label{equation:Ttau}
\begin{split}
    T^{4}(\tau)&=\frac{3}{4}T^{4}_{\rm cool}\bigg[\frac{2}{3}+\tau\bigg]\\
    & +\frac{3T^{4}_{\rm eq}}{4}\bigg[\frac{2}{3}+\frac{1}{\gamma \sqrt{3}}+\bigg(\frac{\gamma}{\sqrt{3}}-\frac{1}{\gamma\sqrt{3}}\bigg)e^{-\gamma \tau \sqrt{3}}\bigg]
\end{split}
\end{align}

\noindent where $\sigma_{\rm SB}T^{4}_{\rm cool}{=}L/4\pi R^{2}_{\rm p}$ is the planet's outgoing flux emitted at the photosphere, $\tau$ is the optical depth (we evaluate equation \ref{equation:Ttau} at $\tau$=2/3), $T_{\rm eq}{=}T_{\star}(R_{\star}/2a)^{1/2}$ is the planetary equilibrium temperature with $a$ the orbital semi-major axis, and $\gamma$ is the ratio of visible to thermal opacities which we take to be 0.4 \cite[appropriate for a hot Jupiter like HD 209658 b;][]{hubbursud03,gui10}.\footnote{Adopting $\gamma{=}0.032$, as appropriate for the super-Earth GJ 1218 b \citep[][]{valguipar13}, changes our planet radii by ${\lesssim}1\%$. For simplicity we do not vary $\gamma$.} 

Equation \ref{equation:Ttau} reflects the $T(\tau)$ solution for planets experiencing isotropic irradiation. Approximating the temperature field as isotropic is appropriate for our tidal migration calculations (Section \ref{sec:ridge}) in which planets spend most of the evolution at large stellocentric distances. We have verified however that using equation 49 of \cite{gui10} for non-uniform irradiation, appropriate for tidally synchronized hot Jupiters/Neptunes, changes our planetary radii by ${<}1\%$.

Throughout this work we assume a Sun-like star, taking $T_{\star}{=}5857$ K and $R_{\star}{=}R_{\odot}$ \citep[values taken from the \texttt{MIST} stellar evolution tracks for a solar-mass star at age 5 Gyr;][]{chodotcon16}. We again use the \texttt{brentq} root finder to iterate over $R_{\rm p}$ until both bottom and top boundary conditions are met.

We do not account for atmospheric layers above the thermal photosphere, unlike some previous works \citep[e.g.][]{jinmorpar14,attbouegg21,haldorven24}. We define the planet radius to be the thermal photosphere, consistent with the definition used by the independent structure codes we benchmark our method against (see Section \ref{sec:mesa}). Our tidal calculations therefore do not depend on the atmospheric structure in the optically thin outer layers.

\subsection{Thermal Evolution $\&$ Grid Construction}\label{subsection:thermal}

We follow planetary thermal evolution by evolving the entropy of the convective interior as the planet cools \citep[``stepping through the adiabats"; e.g.][]{hub77,forhub03,arrbil06}. Because the convective eddy turnover time (${\lesssim}$1 yr; see Appendix \ref{subsection:timescalecheck} for detailed estimates) is much shorter than the global cooling timescale of the planet, we assume that $dS/dt$ is constant throughout the convective zone \citep[e.g.][]{pacsie72}. The entropy equation $\partial L/\partial m{=}{-}T\partial S/\partial t$ then reads \citep[e.g.][]{marcum14},

\begin{equation}\label{equation:cooling}
    \frac{dS}{dt}=\frac{-L}{\int_{\rm conv} T dm}
\end{equation}

\noindent where the integral is over the convective zone. To compute the thermal evolution of a given planet, we look up $L$ and $\int_{\rm conv}Tdm$ from pre-computed tables on the fly as we integrate equation \ref{equation:cooling}. We next detail how we construct our tables. 

Our grid is comprised of five axes along which planets are varied. Each axis and its associated values are summarized in Table \ref{tab:structure_tables}. We choose these five specific axes because each alters the planetary cooling/structure independently of the other axes. Planetary cooling is dictated by the outgoing luminosity $L$, which in turn is set by the radiative convective boundary (RCB). The RCB dictates $L$ because it is the location in the planet where the rate at which energy can be transported outwards reaches a global minimum; convection below the RCB is capable of carrying very large luminosities, while the maximum luminosity that radiative diffusion can carry ($L{\propto}mT^{4}/(\kappa P)$ via equation \ref{equation:nablas}) is minimized at the base of the radiative layer since opacities grow with pressure \citep[][]{arrbil06}. The RCB is therefore the bottleneck for cooling. Each axis in Table \ref{tab:structure_tables} alters the location of the RCB, and therefore must be accounted for.

The first axis we vary is planet mass, from $m_{\rm p}{\in}[10.1,300] \ M_{\oplus}$, with a resolution of $\Delta m_{\rm p}{=}0.1 \ M_{\oplus}$ for atmospheres ${\leq}1 \ M_{\oplus}$ and $\Delta m_{\rm p}{=}2 \ M_{\oplus}$ otherwise (this resolution is a balance between speed and computing the grid without loss of accuracy). Second we explore two core masses $m_{\rm c}{\in}\{10,20\} \ M_{\oplus}$. 
Third we use three atmospheric metal mass fractions $Z$: solar $Z{=}0.02$ (hydrogen mass fraction $X{=}0.74$), moderately super-solar $Z{=}0.1$ ($X{=}0.65$) and very super-solar $Z{=}0.5$ ($X{=}0.36$). The corresponding $X$ values for each $Z$ are chosen to be consistent with the composition of the opacity tables we adopt (detailed in Section \ref{subsec:kappa}); e.g. $Z{=0.5}, \ X{=0.36}$ yield a metal number fraction [M/H]${=}$1.7, in agreement with the opacity table we employ. Fourth we vary the entropy in the innermost convective zone $S$, ranging from ${\sim}6{-}11 \ k_{\rm B}/m_{\rm H}$ for $Z{=}\{0.02,0.1\}$ and ${\sim}4{-}8  \ k_{\rm B}/m_{\rm H}$ for $Z{=}0.5$ ($k_{\rm B}$ is Boltzmann's constant and $m_{\rm H}$ the hydrogen mass). Lastly we vary the equilibrium temperature $T_{\rm eq}{\in}\{2884,2039,1442,912,288\}$ K corresponding to distances 0.01, 0.02, 0.04, 0.1, and 1 au from our Sun-like star respectively (also chosen to balance speed of computing without loss of accuracy in interpolation). This range of equilibrium temperatures brackets the orbital distances traversed by planets we explore in this work and Paper II, so that we interpolate within these boundaries during our dynamical integrations.

For each model of core mass $m_{\rm c}$, total mass $m_{\rm p}$, envelope metallicity $Z$, luminosity $L$, and equilibrium temperature $T_{\rm eq}$, we tabulate the radius, entropy $S$, and the corresponding $\int_{\rm conv} T dm$ using the procedure outlined in Section \ref{subsec:static_planets}. 

\subsection{Extra Heating}

Our formulation also allows us to add extra heating to the thermal evolution. In the case that the extra luminosity $L_{\rm extra}$ is deposited below the RCB \citep[the effect of extra heating emanating from the core or inside the convective zone is identical; e.g.][]{mil19}, the entropy equation includes a term that offsets the cooling \citep[e.g.][]{huacum12}:

\begin{equation}\label{equation:cooling_Lx}
    \frac{dS}{dt}=\frac{-L+L_{\rm extra}}{\int_{\rm conv}T dm}.
\end{equation}

\noindent Equation \ref{equation:cooling_Lx} expresses the fact that $L_{\rm extra}$ decreases the fraction of the luminosity $L$ that cools the interior.

On the other hand, extra heating deposited at a pressure above the convection zone $P_{\rm dep}{<}P_{\rm RCB}$ alters the cooling in a less straightforward manner. In this case, extra heating may change the size, location, and even the number of RCBs \citep[e.g.][]{huacum12,komyou17}. To treat this flavor of heating, we take advantage of the fact that in thermal equilibrium $dL/dr{\geq}0$ throughout the planet. For a given planet model, we therefore introduce a step function $L(r)$ into the structure via:

\begin{align}\label{equation:LPdep}
\begin{split}
    L(r)&=L; \  P(r)>P_{\rm dep} \\
        &=L+L_{\rm extra}; \ P(r)\leq P_{\rm dep},\\
    \mathrm{where} \ P_{\rm dep} &< P_{\rm RCB}   .
\end{split}
\end{align}

\noindent For a given $m_{\rm p}$, $L$, $L_{\rm extra}$ and $P_{\rm dep}$, we solve the planet structure using the same root-finding procedure as without extra heat, taking $S$ and $\int_{\rm conv} T dm$ across the innermost convection zone. Thermal evolution is computed using equation \ref{equation:cooling}. 

\subsection{Equation of State}\label{subsec:eos}

We employ the SCvH equation of state for hydrogen and helium from \cite{saucha95}, which covers temperatures $125{-}10^{7}$ K, and pressures $10^{4}{-}10^{19}$ dyn cm$^{-2}$. We use the Frankfurt equation of state for SiO$_{2}$ \cite[\texttt{FEOS};][]{faitauios18} as a proxy to represent heavy elements.\footnote{ \url{https://data.mendeley.com/datasets/6vjsv6v48p/1}} Using \texttt{FEOS}, we tabulated the SiO2 equation of state across a density and temperature range $5{\times}10^{-11}$ to $50$ g cm$^{-3}$, and 116-$10^{6}$ K, respectively. To combine the individual hydrogen, helium, and SiO$_{2}$ equations of state into one for their mixture, we adopt the linear mixing approximation \citep[augmented with a correction for the entropy of mixing; e.g.][]{saucha95,vazkovpod13}:

\begin{align}
\begin{split}
    \rho(P,T)&=\bigg(\frac{X}{\rho_{\rm H}(P,T)}+\frac{Y}{\rho_{\rm He}(P,T)}+\frac{Z}{\rho_{\rm Z}(P,T)}\bigg)^{-1}\\
    S(P,T)&=XS_{\rm H}(P,T)+YS_{\rm He}(P,T)+ZS_{\rm Z}(P,T)\\
    &+S_{\rm mix}(P,T),
\end{split}
\end{align}

\noindent where $\rho_{\rm H/He/Z}(P,T)$ is the density of hydrogen/helium/metals each evaluated from their equations of state, and $S_{\rm mix}$ is the entropy of mixing. Our calculation of $S_{\rm mix}$ follows equations 50 and 51 from \cite{saucha95}, corrected to include contributions from the mixing of metals with hydrogen and helium (see Section \ref{subsection:timescalecheck} for details on our calculation of $S_{\rm mix}$). We also include contributions to $S_{\rm mix}$ from metal ionization using the charge states tabulated by \texttt{FEOS}. The combined equation of state tables we make follow the same resolution and parameter space as SCvH up to a maximum temperature of $5{\times}10^{5}$ K.

An important limitation of our heavy element EOS is that we do not account for water. Using water as a heavy element proxy rather than SiO$_{2}$ would produce less compression at a given $Z$ owing to the lower density of water than silicate \citep[e.g.][]{vazkovpod13,rog25}. We expect that using water as opposed to silicate would therefore quantitatively alter the results presented in Section \ref{sec:ridge} for metal-enriched planets (see discussion in that section). The qualitative conclusions drawn from this paper are unlikely to change, however.

Once we compute $S(P,T)$, we tabulate $\nabla_{\rm ad}(P,T)$ using finite differences to compute the derivatives in equation \ref{equation:nablas} (cubic splines are used at grid edges). We also tabulate the mean molecular weight $\mu(P,T)$ and adiabatic exponent $\Pi{=}\partial \log P/\partial \log \rho |_{S}$ (to be used in Paper II).

We occasionally find that the photospheric pressure in our planet models lies below the SCvH lower limit of $10^{4}$ dyn cm$^{-2}$. At pressures less than $10^{4}$ dyn cm$^{-2}$, we switch to an ideal gas equation of state via $\rho(P,T){=}P\mu(P,T) m_{\rm H}/k_{\rm B}T$ where $\mu{=}1/(X/\mu_{\rm H}{+}Y/\mu_{\rm He}{+}Z/\mu_{\rm Z})$. We take $\mu_{\rm He}{=}4$, $\mu_{\rm Z}{=}60$ (to yield continuity with our SiO$_{2}$ equation of state), and $\mu_{\rm H}$ from H$_{2}$ dissociation equilibrium. We follow the procedure described in Section 2.1.2 of \cite{leechiorm14} to compute $\mu_{\rm H}$ via the Saha equation using the partition functions of atomic and molecular hydrogen \cite[see also][]{hubmih14}. We have verified that our $\mu$ for solar metalicity gas agree within a few percent with those shown Figure 1 of \cite{koslavhua22}, which were independently calculated using the NASA Chemical Equilibrium Applications Code \citep[][]{gormcb94}.

\subsection{Opacities}\label{subsec:kappa}

We numerically interpolate the Rosseland mean opacities $\kappa_{\rm R}$ (evaluated at local temperature) from \cite{freelusfor14} \cite[which are extensions of the opacities from][]{fremarlod08}. These opacities cover pressures $10^{0}{-}3{\times}10^{8}$ dyn cm$^{-2}$ and temperatures 75${-}$4000 K,  as well as metallicities [M/H]=0${-}$1.7 (i.e. solar to 50${\times}$ solar, by number; $Z/X{=}(Z_{\odot}/X_{\odot})10^{[\rm M/H]}$ in terms of mass fraction). We use the [M/H]${=} 1.7$ tables for our $Z{=} 0.5$ EOS. Electron conduction opacities $\kappa_{\rm c}$ from \cite{potponjos15} are used together with $\kappa_{\rm R}$ to produce the total opacity via $\kappa{=}1/(\kappa^{-1}_{\rm R}{+}\kappa^{-1}_{\rm c})$.\footnote{\url{http://www.ioffe.ru/astro/conduct/condint.html}} For solar composition gas, the average ion charge number is $\bar{\rm Z}{=}\Sigma_{i}z_{i}y_{i}/\Sigma_{i}y_{i}{\sim}1$, where $z_{i}$ and $y_{i}$ are the atomic number and number fraction of each species \citep[e.g.][]{paxbildot11}, so we evaluate $\kappa_{\rm c}$ from the $\bar{\rm Z}{=}1$ table of \cite{potponjos15}. For our $Z{=} 0.5$ mixture we linearly interpolate between the $\bar{\rm Z}{=}1 \ \mathrm{and} \ \bar{\rm Z}{=}2$ tables since $\bar{\rm Z}{\sim}  1.6$. 

\begin{figure}
\epsscale{1.2}
\plotone{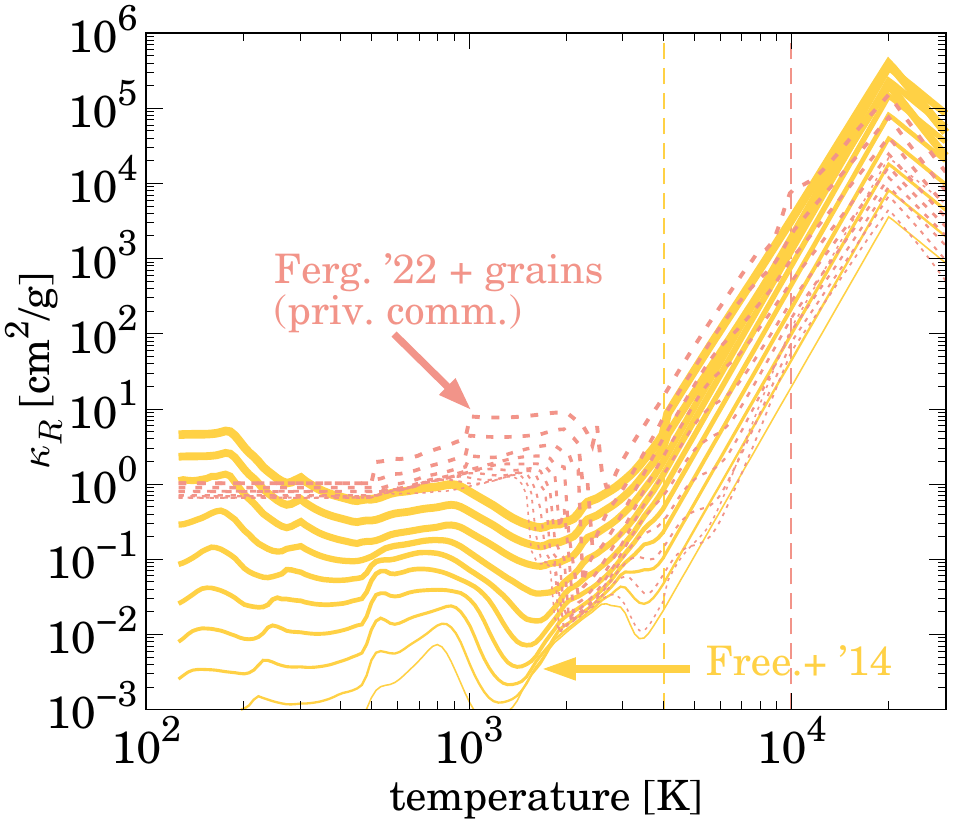}
\caption{Comparison of Rosseland mean opacities as tabulated by \cite{freelusfor14} (``Free.+ '14" in orange) and by J. Ferguson (private communication, ``Ferg. '22 ${+}$ grains" in red). Thin to thick (bottom to top) lines correspond to ten pressure values logarithmically distributed between $10^{4}{-}10^{9}$ dyn cm$^{-2}$. The vertical lines denote the maximum temperatures contained in each opacity table (i.e. data to the right of the lines is extrapolated; the Ferguson tables are also extrapolated below 500 K). At temperatures ${\lesssim}1500$ K, dust grain opacity included in the tables from J. Ferguson sets a lower bound on $\kappa_{\rm R}{\sim}1$ cm$^{2}$ g$^{-1}$, in contrast to the lower molecular opacities contained in \cite{freelusfor14}. The steep rise in $\kappa_{\rm R}$ to $2{\times}10^{4}$ K reflects $H^{-}$, whereas the dropoff at higher temperatures is due to our assumption of bound-bound/bound-free/free-free opacities dominating.\label{figure:kappacomp}}
\end{figure}

We also require pressures ${>}3{\times}10^{8}$ dyn cm$^{-2}$ and temperatures ${>}4000$ K, so for this we extrapolate the opacity tables of \cite{freelusfor14}. At temperatures $4{\times}10^{3}{\lesssim}T{\lesssim}2{\times}10^{4}$ K, opacity is dominated by $H^{-}$ owing to thermal ionization of metals so that $\kappa_{\rm R}{\propto}\rho^{0.5}T^{7.5}$ \citep[e.g.][see also \citealt{leechiorm14}]{kipwei90,vazkovpod13}. At higher temperatures $T{\gtrsim}2{\times}10^{4}$ K, the ionization fraction grows so that free-free, bound-free, and bound-bound transitions dominate the opacity, yielding a Kramer's scaling $\kappa_{\rm R}{\propto}\rho T^{-3.5}$ \citep[][]{vazkovpod13}. To extrapolate to pressure and temperature $(P_{\rm extrap},T_{\rm extrap})$, we first identify the nearest $(P,T)$ point contained in the tables of \cite{freelusfor14}. We then scale this opacity from the table up to $(P_{\rm extrap},T_{\rm extrap})$ following the $H^{-}$ relation and, if $T{>}2{\times}10^{4}$ K, Kramer's relation. We scale with density by mapping $P_{\rm extrap}{\rightarrow}\rho_{\rm extrap}$ using our EOS. Our opacities for solar composition gas are shown in Figure \ref{figure:kappacomp}.

Figure \ref{figure:kappacomp} also displays opacities provided by J. Ferguson in a private communication, which we include for comparison. These tables build on those from \cite{feraleall05} which use the \texttt{PHOENIX} stellar atmosphere code. They cover temperatures $500{-}10^{4}$ K and $\log{R}{=}\log{\rho}{-}3\log{T}{+}18$ from ${-}8$ to $8$.  Importantly, the Ferguson tables allow metals to take the form of dust grains, as opposed to the tables from \cite{freelusfor14} that assume metals are only molecular. We use the same extrapolation procedure as outlined above. For $T{<}500$ K, we assume $\kappa_{\rm R}$ is constant and the same as at 500 K due to the presence of dust grains which set a lower limit on opacity (${\sim}1$ cm$^{2}$ g$^{-1}$). As shown in Figure \ref{figure:kappacomp}, inclusion of grains yields significantly higher opacities at low temperatures. Both tables display reasonable agreement in the regime where dust sublimates and H$^{-}$ takes over however ($2000{<}T{<}10^{4}$ K), confirming that our extrapolation procedure of \cite{freelusfor14} in this regime is correct. We employ the opacities of \cite{freelusfor14} for the fiducial calculations shown in this paper (see Figure \ref{figure:ss_zinflate} however for calculations using grain opacities). Our choice of using the \cite{freelusfor14} opacities as fiducial values has two justifications. First, dust grains are likely to settle out of planetary upper radiative zones over timescales shorter than the ages of evolved planets we consider here \citep[e.g.][]{movpod08}. Second, we seek to benchmark our code against \texttt{MESA}, which does not employ dust opacities.

\begin{figure}
\epsscale{1.26}
\plotone{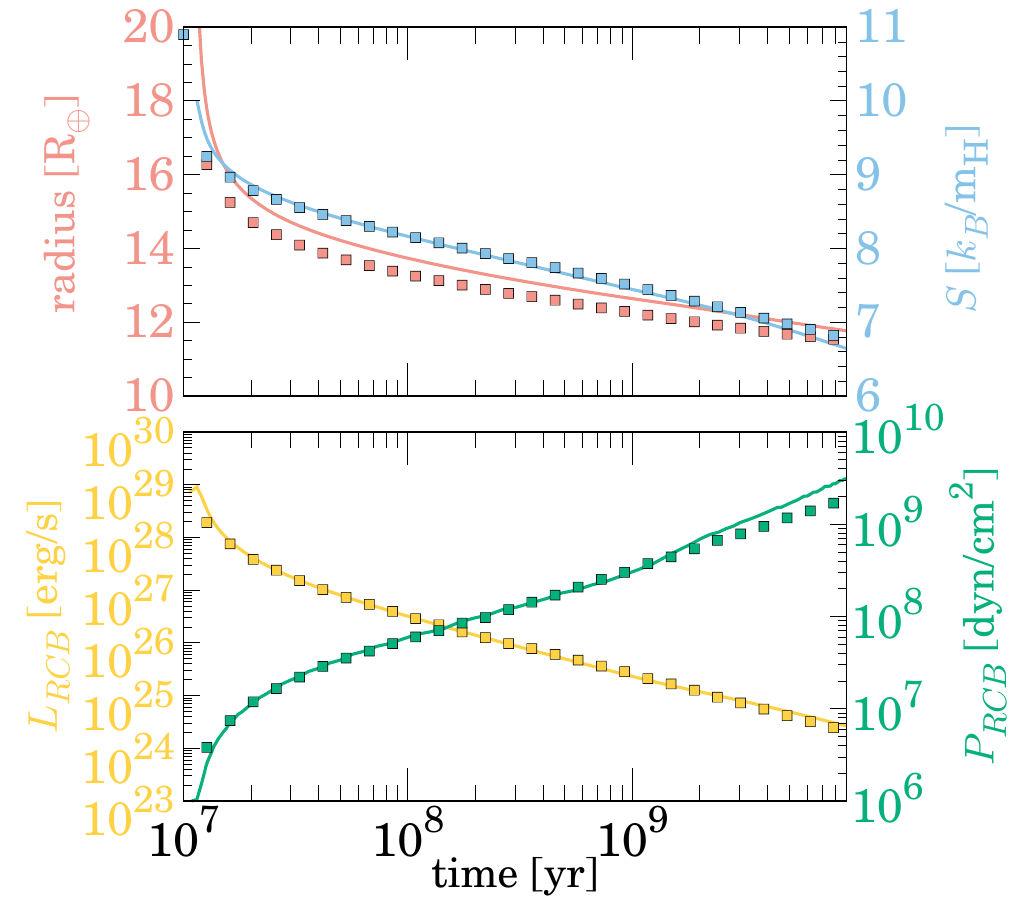}
\caption{Thermal evolution of a fiducial hot Jupiter (300 $M_{\oplus}$, 10 $M_{\oplus}$ core) at an equilibrium temperature of 1440 K computed using \texttt{MESA} (in solid lines) and our own thermal evolution code (in square points; computed using dust-free opacities). While planetary radii are offset by ${\lesssim}3\%$ (red), internal entropy (blue), luminosity at the RCB (orange), and pressure at the RCB (green) agree to within ${\lesssim}1\%$ for the entire evolution. \label{figure:mesa_comp1}}
\end{figure}

\section{Benchmarking Against \texttt{MESA}}\label{sec:mesa}

We begin by benchmarking our thermal evolution model against the open source stellar/planetary evolution code \texttt{MESA} \citep[version 11554;][]{paxbildot11,paxcanarr13,paxmarsch15,paxschbau18,paxsmosch19}. \texttt{MESA} employs custom radiative opacity tables built upon those from \cite{fremarlod08}, and custom conduction opacities built from \cite{caspotpie07}, but uses the same hydrogen/helium SCvH equation of state that we do. Our \texttt{MESA} models employ $Z{=}0.02$ to accord with our own structure calculations (we note that \texttt{MESA} treats metals as ideal gases in the planetary regime, unlike our non-ideal EOS for SiO$_{2}$). Since \texttt{MESA} does not contain a high metallicity equation of state in the planetary regime or high metallicity opacities, we benchmark it against our model only for solar composition planets (see Section \ref{subsec:highZ} for metal-rich benchmarking). We construct irradiated \texttt{MESA} planets using the approach outlined in \cite{hallee22}. As a brief summary: we insert an inert core into the coreless Jupiter-like model from the \texttt{irradiated$\_$planet} test suite of \cite{paxcanarr13}. The model is irradiated via the \texttt{relax$\_$irradiation} routine, in which we distribute the irradiation heating across the outermost column of 500 g cm$^{-2}$ \cite[corresponding to an opacity to visible light of $4{\times}10^{-3} \ \rm cm^{2} \ g^{-1}$;][]{gui10}. We do not include radiogenic heating or thermal inertia from the core. Planet radii are taken where the optical depth to outgoing thermal radiation is $2/3$, in accordance with our own structure calculations. To construct lower mass sub-Saturn/hot Neptune planets ($m_{\rm p}{\sim}10{-}30 \ M_{\oplus}$), we first locate the model from the grid of \cite{hallee22} with mass closest to our desired mass. We then adjust the mass using the \texttt{relax$\_$mass} routine to that desired.

Our benchmarking results are organized as follows. In Section \ref{subsec:HJcooling} we benchmark cooling curves for irradiated hot Jupiters. Section \ref{subsec:HJheating} showcases hot Jupiter evolution under extra heating, while Section \ref{subsec:neptunes} deals with lower mass Neptune-like planets. Finally, Section \ref{subsec:highZ} compares evolution of metal-enriched giant planets and Neptunes with the \texttt{planetsynth} code from \cite{mulhel21}.

\begin{figure}
\epsscale{1.26}
\plotone{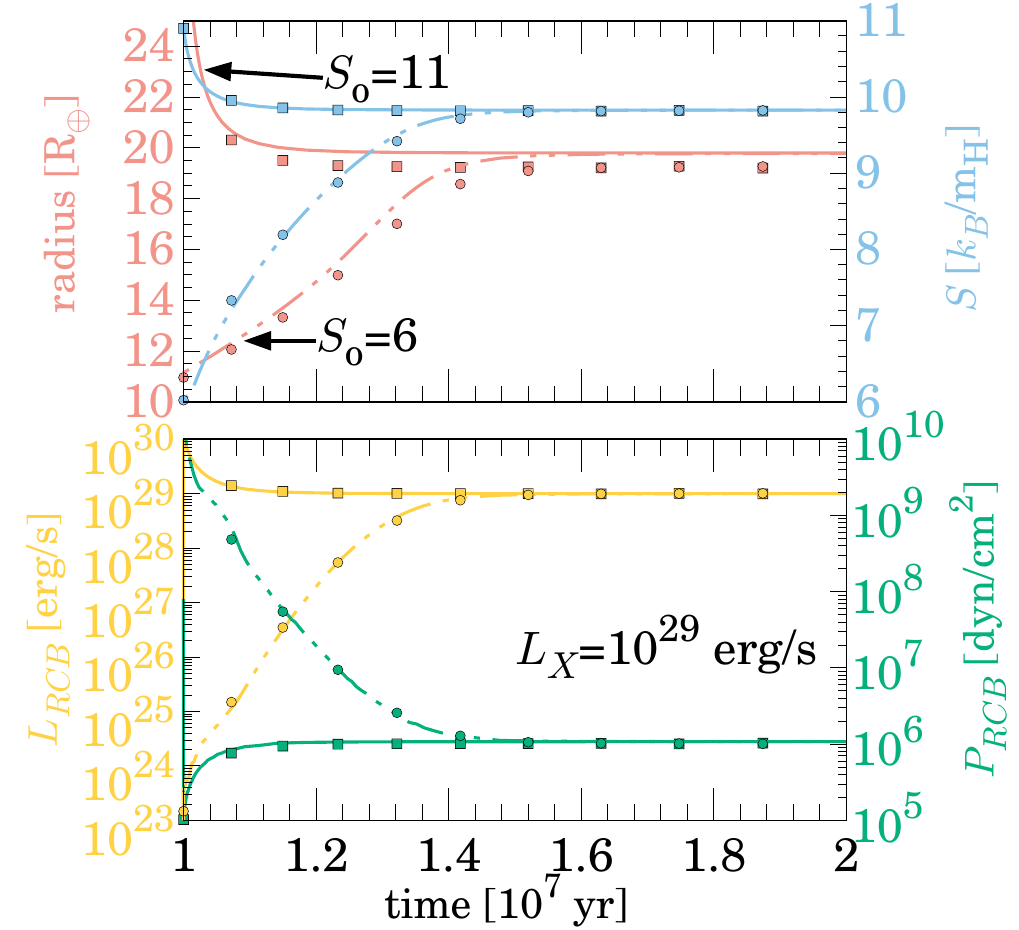}
\caption{Planetary evolution in the presence of $L_{\rm X}{=}10^{29} \ \mathrm{erg \ s^{-1}}$ of extra heat deposited at the base of the convective zone, for the same planet as Figure \ref{figure:mesa_comp1}. Solid curves and square data points display evolution tracks computed with \texttt{MESA} and our own model, respectively, for a planet that begins with a high entropy ($S_{0}{=}11 \ k_{\rm B} / m_{\rm H}$). The planet cools until it reaches a steady state when the extra heat is balanced by radiative losses. Dot-dashed lines accompanied by circular points highlight evolution for a planet with initially low entropy ($S_{0}{=}6 \ k_{\rm B} / m_{\rm H}$). This ``cold start" planet rapidly inflates over a puff-up timescale ${\sim}S_{0}/|dS/dt|_{0}{\sim}10^{5}$ yr until steady state is achieved. Since the puff-up time exceeds the convective turnover time and radiative time of the planet, structural and thermal equilibrium are maintained throughout radius expansion. This equilibrium allows our approach to perform just as well as \texttt{MESA}. \label{figure:mesa_comp2}}
\end{figure}

\subsection{Hot Jupiter Cooling}\label{subsec:HJcooling}

Figure \ref{figure:mesa_comp1} displays thermal evolution for a fiducial hot Jupiter computed with our approach versus \texttt{MESA}. Internal entropy as well as conditions at the RCB agree to within ${\lesssim}1\%$ for the entire evolution. Our planetary radii are offset lower compared to that computed by \texttt{MESA} by ${\sim}3\%$. We have traced this discrepancy in radius to a combination of differing opacity tables (\cite{fremarlod08} vs. \cite{freelusfor14}, for \texttt{MESA} and our approach respectively) as well as differing treatment of irradiation; the \texttt{MESA} model distributes irradiation heating over the outermost column of 250 g cm$^{-2}$, whereas our models employ equation \ref{equation:Ttau}. We find that using shallower heating depths ${\lesssim}100 \ \rm g \ cm^{-2}$ brings the radius into closer agreement within ${\lesssim}1\%$.  

\begin{figure}
\epsscale{1.26}
\plotone{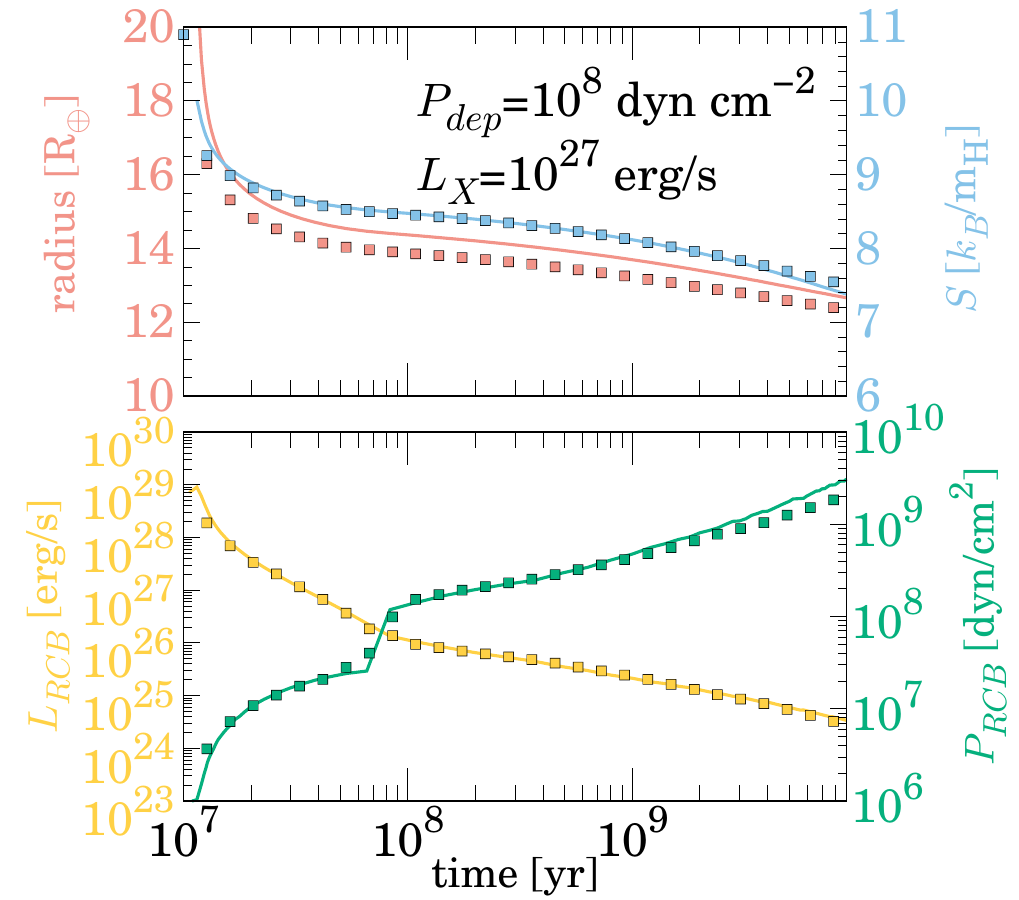}
\caption{Cooling curve for our fiducial hot Jupiter experiencing an additional $10^{27}$ erg/s of extra heat deposited at a pressure $P_{\rm dep}{=}10^{8}$ dyn cm$^{-2}$. The rapid increase in $P_{\rm RCB}$ at ${\sim}7{\times}10^{7}$ yr stems from the opening of a radiative window below the deposition pressure which bottlenecks the cooling for the remainder of the evolution. The luminosity shown in orange is the \textit{cooling} luminosity measured at the innermost RCB. Agreement between our models and \texttt{MESA} is once again excellent. \label{figure:mesa_comp3}}
\end{figure}

\

\subsection{Hot Jupiters Under Extra Heating}\label{subsec:HJheating}

We next compare structure evolution when extra internal heating is included. Figure \ref{figure:mesa_comp2} showcases planets that experience extra heating injected at the base of the planet (i.e. the bottom of the convective zone) with hot and cold initial conditions (i.e. high and low entropy). Regardless of the initial entropy, \texttt{MESA} and our approach again agree to better than ${\sim}1\%$. Such strong agreement is expected since planets that begin at low entropy heat up over a timescale ${\sim}S/|dS/dt|{\sim}10^{5}$ yr (for our chosen extra luminosity) which exceeds both the radiative time ${\sim}10^{4}$ yr and convective turnover time ${\sim}10^{-1}$ yr of the planet model, allowing it to maintain structural and thermal equilibrium as it inflates. Rather than inflating, high entropy planets on the other hand simply cool until they achieve a thermal steady state during which the cooling luminosity balances the extra heat.

\begin{figure}
\epsscale{1.26}
\plotone{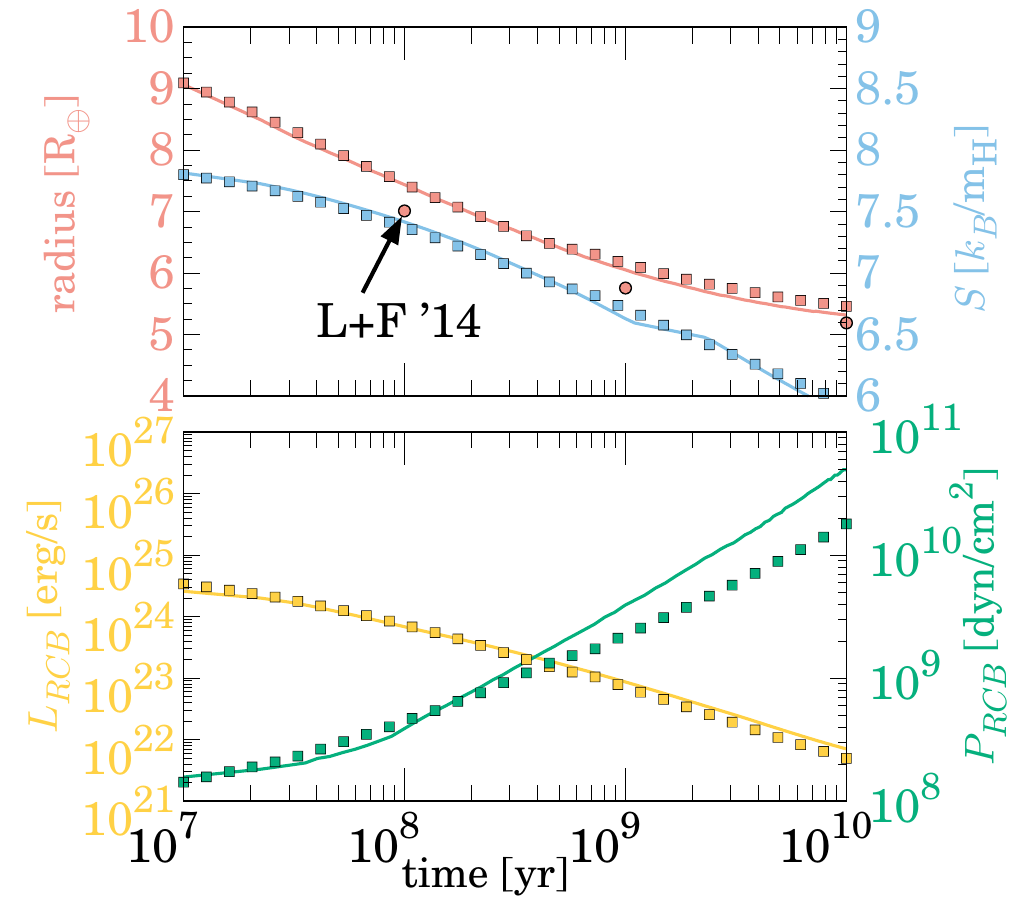}
\caption{Thermal evolution of a hot Neptune (13 $M_{\oplus}$, 10 $M_{\oplus}$ core) at an equilibrium temperature of 1440 K computed using \texttt{MESA} (in solid lines) and our own thermal evolution code (in square points). Planetary radii from the tables of \cite{lopfor14} are also shown in red circular points for comparison (``L+F '14"; for these data points we interpolate their solar metallicity tables for 13 $M_{\oplus}$ planets with envelope mass fractions $20\%$ linearly in irradiation flux to match our planet). Planetary radii, entropy, and RCB properties agree within ${\lesssim}$a few$\%$}. \label{figure:mesa_comp4}
\end{figure}

We next explore in Figure \ref{figure:mesa_comp3} the effect of extra heating deposited at a specific pressure following the luminosity profile in equation \ref{equation:LPdep}. In the specific case that we explore (depositing the heat at pressure $10^{8} \ \rm dyn / cm^{-2}$), the planet cools in two different regimes during the evolution. Early on when the internal cooling luminosity exceeds the extra heat, cooling proceeds as usual with the entropy of the single convective zone dropping. Once the cooling luminosity in the convective zone is sufficiently lower than the extra heating rate, a second, near-isothermal radiative zone opens up below the deposition radius; the drop in $L$ from pressures lower to higher than the deposition pressure outweighs the increase in $P$ to drive $\nabla_{\rm rad}{\propto}LP{<}\nabla_{\rm ad}$ \citep[see Figure 6 of][]{komyou17}. The cooling rate is then dictated by the bottleneck at the innermost RCB, which continues to move to larger pressures as it cools. 

\subsection{Neptune-Like Planets}\label{subsec:neptunes}

We next benchmark our models for lower mass planets. Figure \ref{figure:mesa_comp4} illustrates cooling curves for a hot Neptune-like planet according to our grid, \texttt{MESA}, as well as the planetary radii tables from \cite{lopfor14}. Planetary radii are in excellent agreement (${\lesssim}$a few$\%$) between the three independent methodologies. Figure \ref{figure:mesa_comp4} also indicates a moderate discrepancy in the location of the RCB between our grid and \texttt{MESA} (though this error does not significantly affect the radius, entropy, or luminosity evolution). We have traced this difference to variations in radiative opacity extrapolation methods; \texttt{MESA} extrapolates in the low temperature, high density regime occupied by the planet in Figure \ref{figure:mesa_comp4} by taking the opacity at $\log R{=}1$ ($\log \rho{\sim}{-6}$ for the planet here) before combining with conduction opacities (see \citealt{paxbildot11} and \citealt{paxcanarr13}, which detail the planetary opacities used by \texttt{MESA}). Our extrapolation scheme yields higher radiative opacities at a given pressure and temperature due to the fact that we scale with density (see Section \ref{subsec:kappa}). Our (slightly) higher radiative opacities compared to \texttt{MESA} therefore push the RCB toward lower pressures.

\subsection{High Metallicity Planets}\label{subsec:highZ}

The last comparison we make is against the \texttt{planetsynth} structure/evolution models of \cite{mulhel21} for high metallicity ($Z{=}0.1$) planets.\footnote{\href{Available here}{https://github.com/
tiny-hippo/planetsynth}} The \texttt{planetsynth} code uses \texttt{MESA}, modified to employ a custom high metallicity EOS for a 50/50 water/rock composition and the updated hydrogen/helium EOS of \cite{chamazsou19}, as well as the \cite{freelusfor14} high metallicity opacities. Irradiation is handled in the same manner as our fiducial \texttt{MESA} calculations \citep[see Appendix A3 of][]{mulhel21}. Namely, \texttt{planetsynth} computes planetary radii where the optical depth to outgoing thermal radiation is 2/3. The irradiation heating is deposited across a fixed column of 300 g cm$^{-2}$, corresponding to an assumed opacity to visible light ${\sim}6{\times}10^{-3}$ g cm$^{-2}$ (slightly larger than our choice of $4{\times}10^{-3}$ g cm$^{-2}$). This heating depth adopted by \texttt{planetsynth} is not varied with planet metallicity, similar to our \texttt{MESA} models.

\begin{figure}
\epsscale{1.26}
\plotone{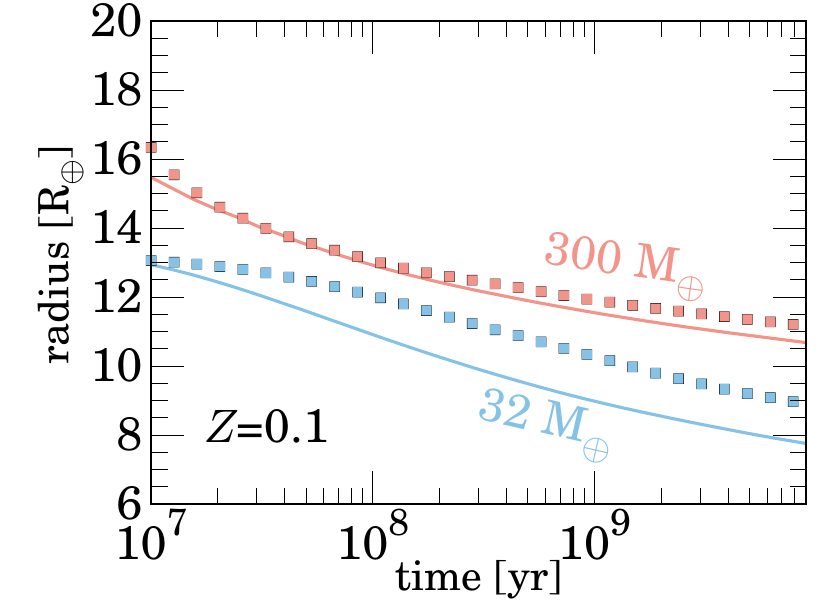}
\caption{Radius evolution for a metal-enriched planet ($Z{=}0.1$) as computed with our approach (square points) versus the \texttt{planetsynth} models of \cite{mulhel21} (solid curves; note that they employ a  different EOS than us). Red and blue curves correspond to 300 and 32 $M_{\oplus}$ planets (with 10 $M_{\oplus}$ cores), respectively, at an equilibrium temperature 288 K. The giant planet's cooling curve is in excellent agreement between the two approaches. Lower mass planets however exhibit ${\sim}15\%$ radius discrepancies, possibly due to EOS differences.} \label{figure:mesa_comp5}
\end{figure}

Figure \ref{figure:mesa_comp5} highlights that, although our giant planet models are again in reasonable agreement, radii for lower mass planets are significantly discrepant (${\gtrsim}10\%$ errors). For the comparison shown in Figure \ref{figure:mesa_comp5}, we chose an initial entropy $S{=}7.6 \ k_{\rm B}/m_{\rm H}$ in order to match the initial radius to the \texttt{planetsynth} model. Despite matching this initial condition, radii remain discrepant over ${\sim}$Gyr. Without detailed knowledge of the planet structure, which is not contained in the \texttt{planetsynth} tables (i.e. entropy and internal luminosity), we cannot identify the source of this discrepancy with certainty. The deviation between our calculations and \texttt{planetsynth} could stem from a difference in hydrogen/helium EOS, since the updated H/He EOS of \cite{chamazsou19} used by \texttt{planetsynth} has been found to produce planets with ${\sim}10\%$ smaller radii than those constructed with SCvH \citep[][]{mulbenhel20,howhelmul25}.

\subsection{Summary of Benchmarking}

We have thus far demonstrated that our approach yields equally accurate results as full stellar/planetary evolution modelling (i.e. using a Henyey code such as \texttt{MESA} to simultaneously solve the partial differential equations of planet structure and energetic evolution). The chief advantage of our method is that it boils down to integrating a single differential equation (equation \ref{equation:cooling}) at run time. The computational speed is therefore limited solely by the efficiency with which the interpolation function is evaluated on the fly. We quantified the calculation speed-up by comparing the time to compute a cooling curve with our approach against \texttt{MESA}, both using default numerical tolerances. We find that our approach is ${\sim}$an order of magnitude faster than \texttt{MESA} (${\sim}1$ second vs. ${\sim}30$ seconds). 

The simplicity of our thermal evolution model also lends itself to straightforward implementation into more sophisticated evolution calculations, as we will show in the next section. For example, interfacing with orbital evolution equations is as easy as calling a \texttt{python} function to supply instantaneous structural parameters of the planet. 

\

\section{Application to Coupled Structure $\&$ Tidal Evolution of Neptunes in the ``Ridge"}\label{sec:ridge}

To illustrate the power of our framework, we now apply our model to study the coupled thermal and tidal evolution of close-in Neptunes at orbital periods ${\gtrsim}3$ days. While Paper II is a dedicated study of hot Neptunes in the ``sub-Jovian desert" (periods ${\lesssim}3$ days), we are motivated here by recent work suggesting that Neptunes immediately outside the desert at periods ${\sim}3{-}6$ days \cite[the Neptunian ``ridge";][]{casboulil24} constitute a distinct sub-population with a unique dynamical history. 

There are several lines of evidence suggesting that Neptunes in the ridge (the subject of this section) possess a unique formation/evolution pathway from those dwelling in the sub-Jovian desert (the topic of Paper II). Neptunes in the ridge exhibit gas-rich compositions \citep[gas mass fractions ${\sim}20{-}40\%$;][see also \cite{caslilarm24}]{doyarmacu25}, in stark contrast to the mostly stripped planets in the desert (see Paper II). Neptunes in the ridge also boast sharply higher occurrence than those in the desert, with the Neptunian orbital period distribution peaking at ${\sim}3{-}6$ days \cite[][see also Figure 1 of \cite{visbeh25}]{casboulil24}. This peak in the orbital period distribution is similar to the peak in hot Jupiter occurrence at ${\sim}3{-}6$ days \citep[e.g.][]{petmarwin18}. The fact that the period distribution resembles that of hot Jupiters, and that many Neptunes in the ridge possess eccentric, misaligned orbits \citep[e.g.][unlike those in the desert that have circular orbits]{corboudes20,bouattmall23}, hint that ridge Neptunes may be emplaced by HEM in a similar manner to hot Jupiters \citep[e.g.][]{boulovbeu18}. This picture also accords with the fact that hot Neptunes inside periods ${\lesssim}6$ days share the same elevated host star metallicity distribution as hot Jupiters, unlike their counterparts at longer periods \citep[][]{visbeh25,doyarmacu25}. In this picture, dynamically hot planets could have undergone recent migration while those on cooler orbits may have had time to tidally damp \citep[][]{doyarmacu25}. In this section, our goal is to use our structure calculations to explore the HEM picture for Neptunes beyond the sub-Jovian desert. We will show that such Neptunes are especially susceptible to runaway tidal inflation, which places strong constraints on the tidal histories of those with low eccentricities.

\subsection{Coupled Orbital-Interior Tidal Evolution}

We consider the case in which a planet on an eccentric orbit circularizes under dissipation in the planet due to tides raised by the star. Because planetary tidal evolution is a very strong function of the instantaneous planetary structure (in the case that tides raised on the planet dominate), the tidal circularization problem is an ideal case study for our model since we can follow the structure self-consistently. We adopt the equilibrium tide model of \cite{hut81}, integrating the following equations of motion due to dissipation in the planet \citep[][]{matpearas10}:

\begin{align}\label{equation:tides}
\begin{split}
\frac{\dot{a}_{\rm tid}}{a}&=-6 k_{2,\rm p}\delta t_{\rm p}\Omega\frac{M_{\star}}{m_{\rm p}}\bigg(\frac{R_{\rm p}}{a}\bigg)^{5}(1-e^{2})^{-15/2}\\
&\times\bigg[f_{1}(e^{2})\Omega-(1-e^{2})^{3/2}f_{2}(e^{2})\omega_{\rm p,PS}\bigg],\\
\frac{\dot{e}_{\rm tid}}{e}&=-27 k_{2,\rm p}\delta t_{\rm p}\Omega\frac{M_{\star}}{m_{\rm p}}\bigg(\frac{R_{\rm p}}{a}\bigg)^{5}(1-e^{2})^{-13/2}\\
&\times\bigg[f_{3}(e^{2})\Omega-\frac{11}{18}(1-e^{2})^{3/2}f_{4}(e^{2})\omega_{\rm p,PS}\bigg],\\
\end{split}
\end{align}

\noindent where $k_{2,\rm p}$ is the planet's Love number (tabulated in our structure grid using the procedure outlined in Appendix \ref{subsection:lovenumber}), $a$ is the semimajor axis, $e$ is the eccentricity, $\omega_{\rm p,PS}$ is the planet's rotation frequency, $\delta t_{\rm p}{=}1/(2\Omega Q_{\rm p})$ is the tidal lag time, with $Q_{\rm p}$ the planet's tidal quality factor \citep[e.g.][]{lai12} and $\Omega{=}(GM_{\star}/a^{3})^{1/2}$ the orbital mean motion with $M_{\star}{=}M_{\odot}$ the stellar mass. We note the steep dependence of $\dot{a}_{\rm tid}$ and $\dot{e}_{\rm tid}$ on the planetary radius ${\propto}R^{5}_{\rm p}$. We further assume that the planet rotates pseudosynchronously at frequency $\omega_{\rm p, PS}$ \citep[e.g.][]{hut81,lecchabar10,matpearas10}:

\begin{equation}
\label{equation:omega_PS}
    \omega_{\rm p, PS}=\frac{f_{2}(e^{2})}{f_{5}(e^{2})}(1-e^{2})^{-3/2}\Omega.
\end{equation}

\noindent Tides raised on the planet by the star dissipate energy at a rate \citep[for pseudosynchronous rotation; e.g.][]{lecchabar10},

\begin{align}\label{equation:Ltid}
\begin{split}
    L_{\rm t}&=3k_{2,\rm p}\delta t_{\rm p}\Omega^{2}\frac{Gm^{2}_{\rm p}}{R_{\rm p}}\bigg(\frac{M_{\star}}{m_{\rm p}}\bigg)^{2}\bigg(\frac{R_{\rm p}}{a}\bigg)^{6}\\
    &\times(1-e^{2})^{-15/2}\bigg[f_{1}(e^{2})-\frac{f^{2}_{2}(e^{2})}{f_{5}(e^{2})}\bigg].
\end{split}
\end{align}

\noindent In this exploration, we assume the tidal luminosity is deposited below the RCB throughout the evolution. In the case that heating is deposited below the RCB, we may use equation \ref{equation:cooling_Lx} with $L_{\rm extra}{=}L_{\rm t}$ to evolve the planetary thermal evolution. Finally, the eccentricity expansion functions $f(e^{2})$ used in equations \ref{equation:tides}--\ref{equation:Ltid} are given by \citep[][]{hut81}:

\begin{align}\label{equation:e_expansion}
\begin{split}
f_{1}(e^{2})&=1+\frac{31}{2}e^{2}+\frac{255}{8}e^{4}+\frac{185}{16}e^{6}+\frac{25}{64}e^{8}\\
f_{2}(e^{2})&=1+\frac{15}{2}e^{2}+\frac{45}{8}e^{4}+\frac{5}{16}e^{6}\\
f_{3}(e^{2})&=1+\frac{15}{4}e^{2}+\frac{15}{8}e^{4}+\frac{5}{64}e^{6}\\
f_{4}(e^{2})&=1+\frac{3}{2}e^{2}+\frac{1}{8}e^{4}\\
f_{5}(e^{2})&=1+3e^{2}+\frac{3}{8}e^{4}.
\end{split}
\end{align}

Altogether, we integrate $\dot{a}_{\rm tid}$ and $\dot{e}_{\rm tid}$ following equations \ref{equation:tides}, and $\dot{S}$ using equation \ref{equation:cooling} with $L_{\rm extra}{=}L_{\rm t}$ from equation \ref{equation:Ltid}. The orbit-averaged stellar flux received by the planet ${\sim}(R_{\odot}/a)^{2}\sigma_{\rm SB}T^{4}_{\star}/(4\sqrt{1-e^{2}})$, yielding an equilibrium temperature $T_{\rm eq}{\sim}T_{\star}(R_{\odot}/2a)^{1/2}/(1-e^{2})^{1/8}$. We look up $R_{\rm p}(S(t),T_{\rm eq}(t))$ on the fly at each time step. We use a fiducial planet mass $m_{\rm p}{=}29 \ M_{\oplus}$ and core mass $m_{\rm c}{=}20 \ M_{\oplus}$, similar to the planets commonly observed in the Neptunian ridge \citep[][]{doyarmacu25}. We discuss how the results change for lower mass planets in Section \ref{subsec:lowmass}.

The planet begins with $a{=}1$ au, $e{=}0.97$, and an initial entropy of $S{=}7 \ k_{\rm B}/m_{\rm H}$. Such a high eccentricity could be achieved via ZLK oscillations excited by an external perturber \citep[e.g.][]{nao16}. In this scenario, $e$ is pumped through ZLK oscillations until short-range forces suppress the oscillations \citep[e.g. general relativistic precession or tides;][]{fabtre07}, at which point isolated tidal decay sets in. This latter phase of post-ZLK tidal circularization is the process we aim to explore here. These initial $a$ and $e$ are chosen such that the planet would end its circularization at an orbital distance $a_{\rm final}{=}a(1-e^{2}){\sim}0.06$ au or an orbital period ${\sim}$5 days, roughly in line with the Neptunian ridge.

In Figure \ref{figure:ss_inflate} we display two sets of circularization evolution tracks. One set uses a constant Love number, while the other varies the Love number in lockstep with the planet structure. For each set we use two slightly different tidal quality factors. Despite the modest differences in the two $Q_{\rm p}$, the evolutionary outcome changes radically between the two cases. This dichotomous behavior is evident in both sets of circularization tracks. In the case with lesser dissipation (larger $Q_{\rm p}$), the planet experiences an early phase of inflation (i.e. $dS/dt{>}0$ since $L_{\rm t}{\gg}L$) before eccentricity damping and planetary cooling catch up ($L{\rightarrow}L_{\rm t}$). The planet then cools and shrinks but never circularizes. A slightly larger amount of dissipation (lower $Q_{\rm p}$) on the other hand yields runaway inflation; the planet radius grows in tandem with tidal heating ($L_{\rm t}{\propto}R^{5}_{\rm p}$) before eccentricity damping or planetary cooling luminosity can slow the inflation down. The decline in Love number as planet radii grow helps counteract tidal heating, but ultimately does not prevent runaway inflation.

\subsection{Analytic Calculation}
\begin{figure*}
\centering
\includegraphics[width=0.9\textwidth]{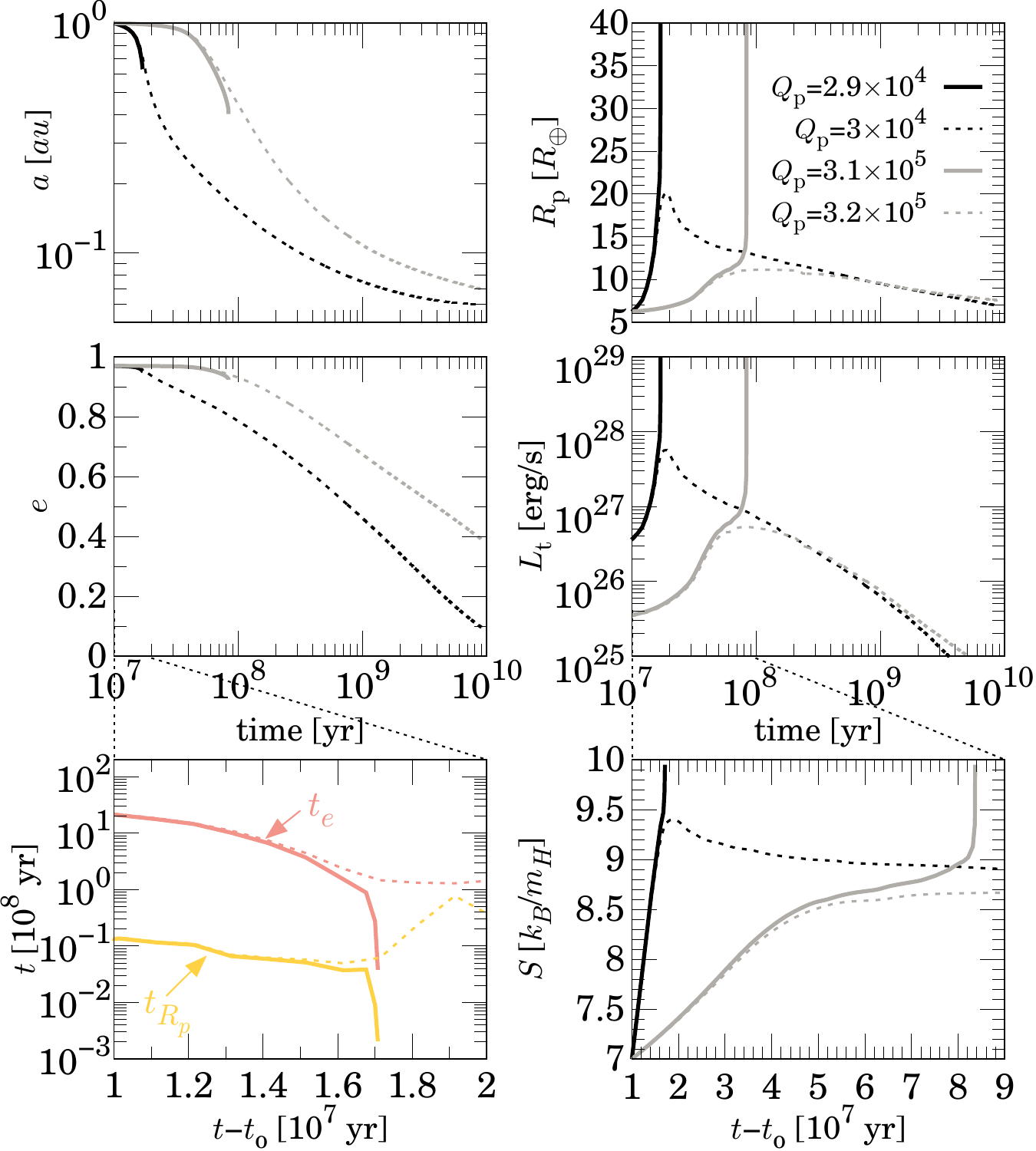}
\caption{Tidal circularization evolution for Neptunian planets around a Sun-like star ($m_{\rm p}{=}29 \ M_{\oplus}$, $m_{\rm c}{=}20 \ M_{\oplus}$, initial $a{=}1$ au, $e{=}0.97$, and $S{=}7 \ k_{\rm B}/m_{\rm H}$). Solid curves correspond to planets that undergo runaway tidal inflation, while dashed curves are stable to tidal inflation. Curves in grey employ a constant Love number whereas curves in black use a Love number that evolves with the planet structure. Semimajor axis ($a$) is shown in the upper left panel, eccentricity ($e$) is shown in middle left, and eccentricity damping time $t_{e}{=}e/|\dot{e}|$ and radius inflation time $t_{R_{p}}{=}R_{\rm p}/|\dot{R}_{\rm p}|$ are displayed in the bottom left (note the zoomed-in $x$ axis; we display $t_{e}$ and $t_{R_{p}}$ only for the planets with varying Love number for clarity). Planetary radii ($R_{\rm p}$) are in upper right, tidal luminosity ($L_{\rm t}$) is in middle right, and entropy ($S$) is shown in bottom right (note again the zoomed-in $x$ axis, which shows $S(t)$ for both sets of planets). Varying the Love number with planet structure stabilizes against tidal inflation and prolongs circularization. Regardless of the Love number evolution, there exists a threshold entropy above which tidal inflation runs away. Planets that avoid runaway inflation do not circularize.
\label{figure:ss_inflate}}
\end{figure*}

The bottom right panel of Figure \ref{figure:ss_inflate} indicates that the transition from tidal damping to runaway inflation during circularization occurs above a threshold entropy $S_{\rm run}$. We now seek to elucidate this transition to runaway inflation (and likely atmospheric destruction) analytically. Our goal is to confirm the numerical result that there is a threshold entropy $S_{\rm run}$ above which runaway is initiated. 

In order to inflate, the tidal luminosity must dominate the radiative cooling luminosity ($L_{\rm t}{>}L$; see equation \ref{equation:cooling_Lx}). In order for inflation to run away, the growth in $L_{\rm t}$ must always outpace that of $L$. Tidal luminosity changes at a rate (equation \ref{equation:Ltid}),

\begin{equation}\label{equation:dLtdR}
    \frac{d\log L_{\rm t}}{d\log R_{\rm p}}=5+\beta,
\end{equation}

\noindent where we have included the dependency $k_{2,\rm p}{\propto}R^{\beta}_{\rm p}$ with $\beta{<}0$. Meanwhile the radiative cooling luminosity increases at a rate,

\begin{align}
\begin{split}
    \frac{d\log L}{d\log R_{\rm p}}&=\frac{d\log L}{d\log S}\frac{d\log S}{d\log R_{\rm p}}.\\
\end{split}
\end{align}

\begin{figure}
\epsscale{1.26}
\plotone{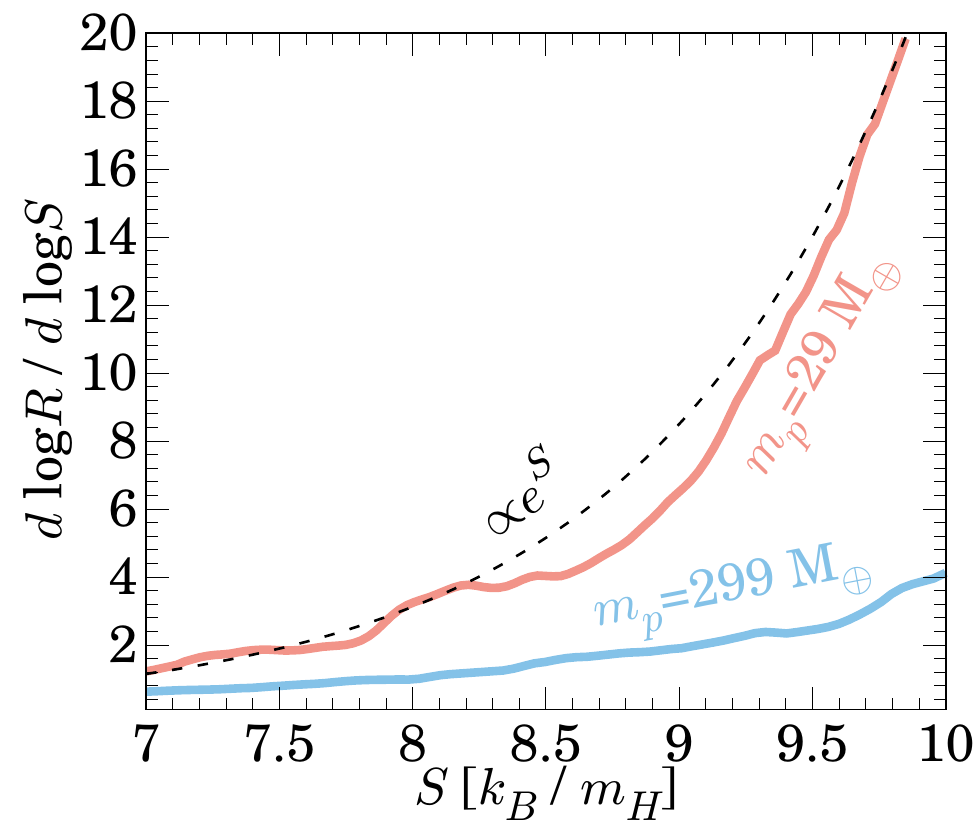}
\caption{Logarithmic derivative of planetary radius with respect to entropy $d\log R_{\rm p}/d \log S$ versus entropy. Colored curves are evaluated directly from our planet structure grids (and lightly smoothed using a Savitzky-Golay filter to aid visualization in this plot). Red and blue curves correspond to planets of masses 29 and 299 $M_{\oplus}$ respectively, both using 20 $M_{\oplus}$ cores and an equilibrium temperature of 288 K. The black dashed curve illustrates that $d\log R_{\rm p}/d \log S$ for the sub-Saturn planet is approximately exponential following ${\sim}1.15 \exp{(S-7)}$. Tidal inflation runaway of sub-Saturns/sub-Neptunes stems from the exponentially growing logarithmic radial response to increasing entropy. \label{figure:gamma}}
\end{figure}

\noindent Following equation \ref{equation:nablas}, the luminosity at the RCB reads
\begin{equation}\label{equation:lrad}
    L=\frac{64\pi Gm\sigma_{\rm SB}T^{4}\nabla_{\rm ad}}{3\kappa P},
\end{equation}

\noindent where we have set $\nabla_{\rm rad}{=}\nabla_{\rm ad}$ reflecting the fact that $L$ is dictated by conditions at the RCB. Using the thermodynamic identity $dS/c_{\rm p}=dT/T-\nabla_{\rm ad}dP/P$, it can be shown that the RCB pressure varies with internal entropy via \citep[see e.g.][and ignoring changes in temperature which are subdominant to pressure]{arrbil06},

\begin{equation}
    P(S)\sim P_{0}\exp\bigg[\frac{-(S-S_{0})}{c_{p}\nabla_{\rm ad}}\bigg],
\end{equation}

\noindent where $P_{0}$ is the RCB pressure at reference entropy $S_{0}$. Writing $\kappa{=}\kappa_{0}(P/P_{0})^{\xi}$ and plugging these expressions for $P(S)$ and $\kappa(P)$ into equation \ref{equation:lrad}, we have (with $S$ is in units of $k_{\rm B}/m_{\rm H}$)

\begin{align}
\begin{split}
    \frac{d\log L}{d\log S}=\frac{k_{\rm B}}{m_{\rm H}}\frac{1+\xi}{c_{p}\nabla_{\rm ad}}S\approx 3.5 S,\\
\end{split}
\end{align}

\noindent where we have used the fact that we find $\xi{\sim}0.5$, likely reflecting $H^{-}$ opacity, and that $c_{\rm p}\nabla_{\rm ad}{\sim}k_{\rm B}/(\mu m_{\rm H})$ with $\mu{\sim}2.33$ evaluated from our EOS. We find empirically that the remaining term $d\log R_{\rm p}/d\log S$ grows exponentially with $S$ following ${\sim}A\exp{[B(S-S_{0})}]$ (see Figure \ref{figure:gamma}) so that in total,

\begin{equation}\label{equation:dLdR}
    \frac{d\log L}{d\log R_{\rm p}}=\frac{3.5S}{A}\exp{[-B(S-S_{0})}],
\end{equation}

\noindent where from Figure \ref{figure:gamma} we estimate $A{=}1.15$ and $B{=}1$ for $S_{0}{=}7$. Equation \ref{equation:dLdR} indicates that the planet's cooling luminosity grows rapidly with increasing radius at low entropy, but grows slowly at large entropy. In contrast, the tidal luminosity always grows rapidly with radius. This leveling off of the radiative cooling luminosity while the tidal heating rate continues to grow produces runaway inflation.

We can now solve for the threshold entropy $S_{\rm run}$ above which tidal inflation enters runaway by setting equations \ref{equation:dLtdR} and \ref{equation:dLdR} equal. This procedure yields,

\begin{align}\label{equation:Sthresh}
\begin{split}
    S_{\rm run}&=\frac{W_{-1}(-(5+\beta)BAe^{-BS_{0}}/3.5)}{B}\\
    &\sim 8.66 \ k_{\rm B}/m_{\rm H}; \ \beta=0 \\
    &\sim 9.24 \ k_{\rm B}/m_{\rm H}; \ \beta=-2\\
\end{split}
\end{align}

\noindent where $W_{-1}$ is the $k{=}{-}1$ branch of the Lambert W function \citep[][]{valjefcor00}. We have evaluated equation \ref{equation:Sthresh} for the case of constant and varying Love numbers ($\beta{=}0,{-}2$ respectively), where the power law index $\beta{=}{-}2$ is an approximate fit to our numerical results (for the planets shown in Figure \ref{figure:ss_inflate}). Equation \ref{equation:Sthresh} shows that a larger entropy is required to achieve runaway when including a structure-dependent Love number due to the decrease in tidal heating rate, in agreement with our numerical calculation. The threshold entropies $S_{\rm run}{\sim}8.66 \ \mathrm{and} \ {\sim}9.24 \ k_{\rm B}/m_{\rm H}$ are also in reasonable agreement with Figure \ref{figure:ss_inflate}.\footnote{We find similar threshold entropies and corresponding $Q_{\rm p}$ when considering circularization of closer-in planets (initial $a{=}0.3$ au and $e{=}0.896$).}

We note that the threshold entropy $S_{\rm run}$ derived above will vary with planet structural parameters that determine the radius/entropy relationship through $A$ and $B$ (i.e. core and total mass, atmospheric metallicity, equilibrium temperature, etc.). We have verified that $d\log R_{\rm p}/d\log S$ does continue to behave exponentially (${\propto} e^{S}$) as we vary each of these axes, though the numerical value of $S_{\rm run}$ does change.

\subsection{Runaway Tidal Inflation for Planets of Varying Mass $\&$ Composition}\label{subsec:lowmass}

\begin{figure*}
\centering
\includegraphics[width=1\textwidth]{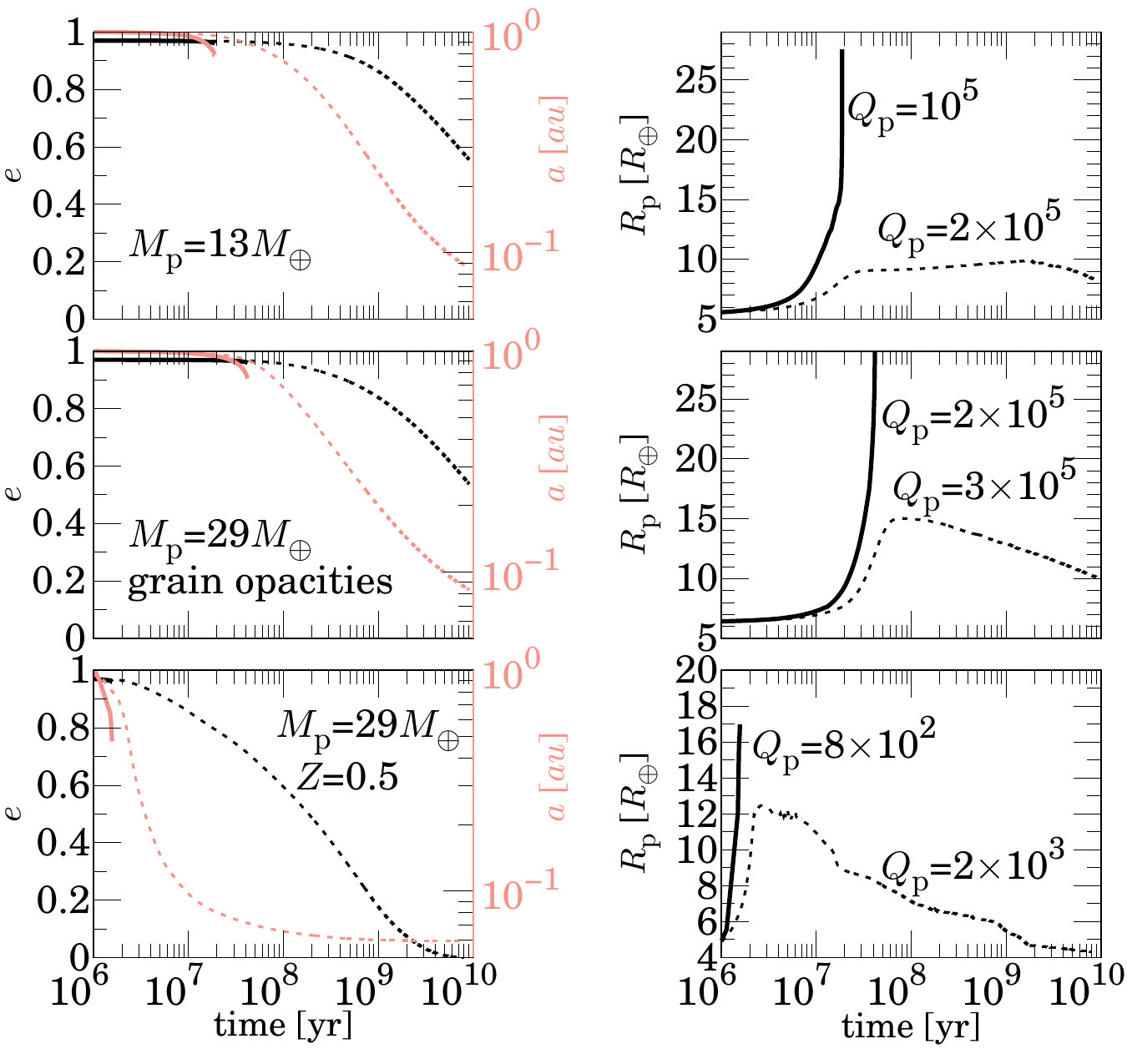}
\caption{Tidal circularization tracks similar to Figure \ref{figure:ss_inflate}, further exploring the effect of planet mass and composition on the prevalence of runaway inflation. Left panels indicate eccentricity (black, left $y$ axis) and semimajor axis (red, right $y$ axis) evolution, while right panels depict planet radius. Each set of left/right panels shows evolution for two $Q_{\rm p}$ values that delineate planets stable and unstable to runaway inflation (dashed and solid curves, respectively). The top panels illustrate that lower mass planets are more susceptible to runaway inflation than higher mass planets (we use 13 $M_{\oplus}$ with a 10 $M_{\oplus}$ core, as opposed to the 29 $M_{\oplus}$ planet in Figure \ref{figure:ss_inflate}). The middle panels highlight that planets with photospheric opacities dominated by dust grains are also more susceptible to runaway inflation than those for which molecular opacities dominate (i.e. those in Figure \ref{figure:ss_inflate}); dust grains retard planet cooling, which cannot keep pace with tidal heating. The bottom panels tell us that planets with high atmospheric metallicities \textit{can} successfully circularize without undergoing runaway inflation (but only for a limited range of $Q_{\rm p}{\in}[2{\times}10^{3},6{\times}10^{3}]$).
\label{figure:ss_zinflate}.}
\end{figure*}

We next explore the coupled tidal and structural evolution of planets with varying mass and composition. Figure \ref{figure:ss_zinflate} displays similar evolution tracks to Figure \ref{figure:ss_inflate} (using the same initial orbital configuration). The upshot of the top panels of Figure \ref{figure:ss_zinflate} is that runaway tidal inflation is exacerbated for lower mass planets owing to their shallower potential wells and correspondingly puffier structures. Dissipation rates that are low enough (i.e. high enough $Q_{\rm p}$) to stave off runaway tidal inflation cannot damp eccentricities significantly; $e{\sim}$0.5 even after 10 Gyr of damping. The minimum eccentricity achieved during circularization without runaway inflation therefore grows towards lower planet masses.

The middle panels of Figure \ref{figure:ss_zinflate} showcase the effect of dust grains in the planetary photosphere on the circularization process. For these runs we employ the opacity tables of J. Ferguson (private communication) that include contributions from grains (see Figure \ref{figure:kappacomp}). We find that dust grain opacities lower planetary cooling luminosities by ${\gtrsim}10{\times}$ at a given entropy; dusty planets cool sluggishly. As a result of this slowed cooling, tidal heating outpaces planet cooling more easily for dusty planets. Dusty planets are therefore more susceptible to runaway inflation than those for which photospheric opacities are dominated solely by molecular sources.

Finally, the bottom panels of Figure \ref{figure:ss_zinflate} indicate that planets with high atmospheric metallicity \textit{can} circularize without undergoing runaway inflation. Metal-rich planets' lower radii (due to the higher density of their envelope constituents) stabilize the planetary puff-up response to tides such that only very high tidal luminosities are capable of driving runaway inflation. High-$Z$ planets still however only circularize without hitting runaway inflation for a limited range of $Q_{\rm p}{\in}[2{\times}10^{3},6{\times}10^{3}]$ (above $Q_{\rm p}{\gtrsim}6{\times}10^{3}$, $e{\gtrsim}0.05$ even after 10 Gyr). Such a finely-tuned range of $Q_{\rm p}$ therefore indicates that runaway tidal inflation remains a threat even to highly metal-enriched planets ($Z{=}0.5$ in Figure \ref{figure:ss_zinflate}). We note that using water instead of SiO$_{2}$ for a heavy element proxy in our EOS would make our metal-rich planets \textit{more} susceptible to runaway inflation at a given $Z$, since water is less dense than silicate.

\subsection{Implications for Close-In Neptune Formation}
The lesson of Figures \ref{figure:ss_inflate} and \ref{figure:ss_zinflate} is that 
forming a low eccentricity Neptune in the ridge via HEM without triggering runaway inflation requires a supersolar planetary metallicity as well as a finely-tuned tidal quality factor $Q_{\rm p}$ (under the assumption that tidal heating is deposited below the RCB). Relatively massive Neptunes with solar composition envelopes (${\sim}30  \ M_{\oplus}$; Figure \ref{figure:ss_inflate}) cannot circularize without triggering runaway inflation. Lower mass planets with solar composition atmospheres (${\sim}10 \ M_{\oplus}$), and Neptunes with photospheric opacities dominated by dust grains (Figure \ref{figure:ss_zinflate}), are even more susceptible to runaway inflation such that they cannot achieve $e{\lesssim}0.5$ without suffering runaway. While we do not attempt to track the post-runaway evolution, the vast majority of the atmosphere is likely to be lost via hydrodynamic escape (e.g. Roche lobe overflow or photoevaporation; see Paper II).

We are therefore left with two possibilities to reconcile low-$e$ Neptunes in the ridge with HEM: either they must exhibit strongly supersolar bulk metallicities and conform to a narrow range of quality factors ($Q_{\rm p}{\in}[2{\times}10^{3},6{\times}10^{3}]$ for atmospheric $Z{=}0.5$), or their tidal dissipation mechanism does not operate below the RCB. In the case that tidal heating takes place above the RCB, much of the extra heat can be bled off into space without inflating the planet (see Figure \ref{figure:mesa_comp3}), while still allowing the orbit to circularize.\footnote{Extremely high tidal luminosities $L_{\rm t}{\gtrsim}10^{30}$ erg/s could still potentially launch a planetary outflow, even if deposited in planetary upper layers. For a typical Neptunian planet with mass and radius $m_{\rm p}{\sim}30 \ M_{\oplus}, \ R_{\rm p}{\sim}6 \ R_{\oplus}$, gas in the photosphere will be heated to a maximum temperature ${\sim}[L_{\rm t}/(4\pi R^{2}_{\rm p}\sigma_{\rm SB})]^{1/4}{\sim}3{\times}10^{3}$ K so that the sonic point $Gm_{\rm p}/2c^{2}_{\rm s}{\sim}40 \ R_{\oplus}{\lesssim}10R_{\rm p}$, making a Parker wind outflow efficient \citep[][]{owewu16}. Future work should address whether such high $L_{\rm t}$ can be achieved in realistic orbital/tidal configurations for close-in Neptunes.}

There are several planets in the Neptunian ridge with circular orbits around FGK stars to which this constraint applies (we focus on circular orbits to make the distinction between runaway inflation and circularization as clear-cut as possible; we also exclude planets with close companions as they are unlikely to arrive via HEM): HATS 7 b \citep[$e{<}0.12$;][]{bondesben17}, HATS 37 b \citep[$e{<}0.345$;][]{jorbakbay20}, HATS 38 b \citep[][]{espstegud24}, WASP 166 b \citep[$e$ consistent with zero;][]{helandtri19}, WASP 156 b \citep[$e$ consistent with zero;][]{pollubbea24}, HD 219666 b \citep[$e{=}0.05{\pm}0.04$;][]{murbeawel25}, TOI 829.01 \citep[$e{=}0$; see][and reference therein]{doyarmacu25}, TOI 4461 \citep[$e{=}0$;][]{doyarmacu25}, TOI 2227.01 \citep[$e{=}0$;][]{doyarmacu25}, and LP 714-47 b \citep[$e{=}0.04{\pm}0.02$;][]{drecrokos20}. If independent lines of evidence are found to suggest that these planets underwent HEM and are not strongly supersolar in bulk metallicity, we therefore must conclude that their tidal dissipation mechanism does not operate below the RCB; e.g. it cannot be dominated by friction in the core \citep[e.g.][]{der79} or turbulent dissipation in the convective zone \citep[e.g.][]{goooh97,ogilin04}. Gravity waves excited in radiative upper layers may be a good candidate \citep[][]{golnic89,lubtouliv97,ogi14}.


To constrain the tidal dissipation mechanism operating within hot Neptunes, follow-up work is therefore needed along two fronts. Firstly it must be determined whether HEM is consistent with hot Neptunes' atmospheric makeup and orbital architectures. Whether Neptunes in the ridge have undergone HEM could be independently constrained via atmospheric spectroscopy with the \textit{James \ Webb \ Space \ Telescope} \citep[JWST;][]{grelinmon16}. Similar to hot Jupiters, elemental abundance ratios (e.g. carbon to oxygen) in close-in Neptune atmospheres may reflect the location at which they accreted gas in protoplanetary disks due to freeze-out of chemical species at differing locations in the disk \citep[e.g.][]{obemurber11}. Future theoretical \citep[e.g.][]{penbookir24} and observational \citep[e.g.][]{ahrganald25} work is needed to tie ridge Neptunes' atmospheric constituents to their formation locations. Measuring the occurrence of distant companions (stellar or planetary) capable of inciting high eccentricities would also help determine the prevalence of HEM in close-in Neptune systems. Determining the occurrence rate of hot Neptunes with nearby planetary companions, and that are therefore unlikely to have arrived via HEM, would also be helpful.

The second axis along which future work is needed is to characterize hot Neptunes' bulk metallicities. Constraining hot Neptunes' metallicities is important because highly supersolar Neptunes remain consistent with HEM (albeit with a finely-tuned $Q_{\rm p}$) and therefore cannot be leveraged to constrain the tidal physics at work in their interiors. Atmospheric spectroscopy with JWST may provide a direct constraint on hot Neptunes' metallicities. Future observations with the Ariel mission \cite[][]{tinecclue22} may help supplement JWST metallicity measurements. Internal structure modeling in tandem with precise mass/radius measurements would also complement direct atmospheric characterization. Such measurements may be facilitated by the PLATO mission \citep[][]{rauaercab25}, in addition to ongoing follow-up campaigns of known hot Neptunes \citep[e.g.][]{ramjensed25}.


Based on Figure \ref{figure:ss_inflate} and the top panels of Figure \ref{figure:ss_zinflate}, we can also place limits on $Q_{\rm p}$ for eccentric planets that avoid triggering runaway inflation (assuming for simplicity solar composition atmospheres). If heating is deposited below the RCB, planets with masses ${\sim}30 \ M_{\oplus}$ must possess quality factors $Q_{\rm p}{\gtrsim}3{\times}10^{4}$ to avoid runaway inflation (modified quality factors $Q^{'}_{\rm p}{=}3Q_{\rm p}/2k_{2,\rm p}{\gtrsim}10^{5}$, similar to Uranus \citep[e.g.][]{titwis90} and Neptune \citep[e.g.][]{banmur92}). For lower mass Neptunes ${\sim}10 \ M_{\oplus}$, these estimates change to $Q_{\rm p}{\gtrsim}10^{5}$ to avoid runaway inflation (modified quality factors $Q^{'}_{\rm p}{\gtrsim}10^{6}$). If heating is deposited above the RCB however, $Q_{\rm p}$ is only constrained by the system age. Neptunes in the ridge that do not harbor substantial gas envelopes, such as TOI 5800 b \citep[with envelope mass fraction ${\lesssim}0.024\%$;][]{napvisbon25,jensydset25}, could also have undergone/are undergoing HEM but could have suffered runaway tidal inflation in the process. 

\section{Future Work}

We will close this paper with a brief discussion of pressing issues for future work tying planetary structure to dynamical evolution. The two major assumptions of our planet structure calculations are 1) that planets possess convective zones in their deep interiors, and 2) that planetary cores are inert and lie below their overlying gas envelopes. In reality, planetary interiors may not be fully adiabatic \citep[see e.g.][for discussion of giant planet interiors]{wahhubmil17,manful21}. Moreover, compositional gradients or mixing processes may blur the distinction between planetary ``core" and overlying ``atmosphere" \citep[e.g.][]{wahhubmil17,vazhelgui18,helste24}. Accounting for how non-adiabatic and diffuse internal structures alter planetary cooling and response to tidal heating could be an important improvement to our calculations. Implementation of such physics into one dimensional planetary evolution codes remains challenging however \citep[][]{mulhelcum20,sursutej24}, and moreover likely requires novel miscibility curves to be calculated for materials in gas-rich planet interiors \citep[e.g. silicate and hydrogen, or hydrogen and water;][]{marguiste22,gupstisch25}.

Future work should also determine whether more realistic core physics may impact planets' structural response to tidal heating. In particular, exchange between superheated cores and overlying atmospheres may alter the core structure in non-trivial ways throughout planets' evolution, depending on e.g. the abundance of water in the envelope which can dissolve/exsolve into/out of molten rock \citep[][]{dorlic21}, and the state of core material which dictates the physics of mixing \citep[i.e. magma or supercritical fluid;][]{marguiste22}. We do not expect however including core heating from radioactive decay or thermal inertia \citep[e.g.][]{lopformil12} to qualitatively change the evolution tracks we compute here as both sources of heat are dwarfed by tidal luminosity during HEM.

Future work should also address whether tidal dissipation via gravity mode excitation in planetary upper layers can indeed circularize planetary orbits without causing radius inflation/mass loss, and if so, what observational signatures may betray the imprint of gravity wave dissipation. We plan to answer this question in an upcoming paper. 

The fate of a planet post-runaway inflation also remains to be clarified (is the entire atmosphere lost?). It is also unclear how different forms of mass loss during inflated migration (e.g. photoevaporation/Roche lobe overflow/inflation-induced boil-off; see \cite{hallee22} and Paper II) help or hinder planets' resilience against further runaway inflation.

\section{Conclusion}\label{sec:conclusion}

In this first of a series of two papers we have showcased and benchmarked a new planet structure code, designed for use in problems that couple planetary interiors and dynamics. We have shown that our calculations compare well with full planetary structure simulations from \texttt{MESA}, for planets evolving under secular cooling as well as under extra heating (e.g. from tides). The chief advantage of our method is its speed and lightweight design: structure calculations run ${\sim}$an order of magnitude faster than planetary evolution models computed using \texttt{MESA}, and interfacing the structure calculations with dynamical integrations is as simple as calling a \texttt{python} function at run time. Our calculation also includes an approximate treatment of metal-enriched planets, a major limitation of the present publicly available version of \texttt{MESA} (our planet structure grid covers metal mass fractions $Z{\in}{\{0.02,0.1,0.5\}}$, assumes heavy elements take the form of silicate, and uses the metal-enriched opacities from \cite{freelusfor14}). We have made the method and planet structure grids we have computed available to the community upon publication of this series of papers (see Footnote \ref{footnote:git}).

We then applied our method to explore the tidal evolution of close-in Neptunes in the newly identified ``ridge" (periods ${\sim}3{-}5$ days), a feature that recent work suggests may be populated via HEM \citep[e.g.][]{bouattmall23,casboulil24,doyarmacu25}. We have shown that there are only two avenues by which low eccentricity Neptunes in the ``ridge" can be emplaced by HEM: 
either they possess highly supersolar metallicities and a narrow range of tidal quality factors ($Q_{\rm p}{\in}[2{\times}10^{3},6{\times}10^{3}]$ for atmospheric metallicity $Z{=0.5}$), or tidal heating cannot be deposited below the RCB, to avoid runaway tidal inflation and subsequent atmospheric destruction. If independent lines of evidence indicate that circularized Neptunes in the ridge indeed arrived via HEM, and that they are not highly metal-rich, our calculations imply that dissipation must occur in the planetary upper layers \citep[e.g. gravity modes excited in radiative zones;][]{lubtouliv97}, rather than e.g. friction in the core \citep[][]{der79} or turbulent dissipation in inner convective zones \citep[e.g.][]{ogilin04}. A promising avenue to independently constrain close-in Neptunes' migration histories and metallicities is to characterize their atmospheres using JWST, similar to ongoing efforts for hot Jupiters \citep[e.g.][]{kirahrpen24,penbookir24}. Determining the fraction of Neptunes in the ridge with distant and nearby companions would also be beneficial to determine whether they underwent HEM, and thereby shed light on the tidal dissipation mechanism operating in their interiors. 

\appendix{}

\section{Stepping through the adiabats: timescale comparison}\label{subsection:timescalecheck}

In order to use our stepping through the adiabats approach to model thermal evolution, both the convective turnover time $t_{\rm conv}$ and radiative time in the radiative envelope $t_{\rm rad}$ must be shorter than the bulk evolutionary time (e.g. cooling, heating, or mass loss) so that the whole planet structure has time to adjust. We explicitly verify below that this requirement is satisfied for a fiducial giant planet and sub-Saturn.

We estimate the convective turnover time using mixing length theory (MLT). Under MLT, the turnover time reads \citep[e.g.][]{coxgiu68,hankaw94},

\begin{equation}
    t_{\rm conv}\sim\bigg(\frac{H}{Q_{\rm exp}g}\frac{1}{\nabla - \nabla_{\rm ad}}\bigg)^{1/2}
\end{equation}

\noindent with $H{=}c^{2}_{s}/g$ the pressure scale height, $Q_{\rm exp}{=}{-}\partial \log \rho /  \partial \log T\big|_{\rm P}$ the thermal expansion coefficient (tabulated in our EOS), and $\nabla{-}\nabla_{\rm ad}$ the superadiabatic temperature gradient. We use scipy's \texttt{brentq} algorithm to numerically solve for $\nabla$ such that the model luminosity $L$ satisfies,

\begin{align}
\begin{split}
    L_{\rm MLT}&=4\pi r^{2} \rho(r) c_{\rm P}T\alpha^{2}_{\rm MLT}H\bigg[\sqrt{\frac{Q_{\rm exp}g(\nabla-\nabla_{\rm ad})^{3}}{H}}\\
    &+\frac{\nu_{T}\nabla}{(\alpha_{\rm MLT}H)^{2}}\bigg],
\end{split}
\end{align}

\noindent where $c_{\rm P}$ is the specific heat capacity at constant pressure (tabulated in our EOS), $\alpha_{\rm MLT}{=}1$, and $\nu_{\rm T}{=}16\sigma_{\rm SB}T^{3}/3\kappa\rho^{2}c_{\rm P}$. We find typical values $\nabla{-}\nabla_{\rm ad}{\sim}10^{-9}{-}10^{-8}$, yielding $t_{\rm conv}{\sim}$days${-}$weeks.

The radiative time is given by,

\begin{equation}
    t_{\rm rad}{\sim}\frac{4\pi}{L}\int^{\rm ph}_{\rm RCB} c_{\rm p}(r)T(r) r^{2}\rho(r)dr,
\end{equation}

\noindent where the integral extends from the RCB to the photosphere.

\begin{figure}
\epsscale{0.6}
\plotone{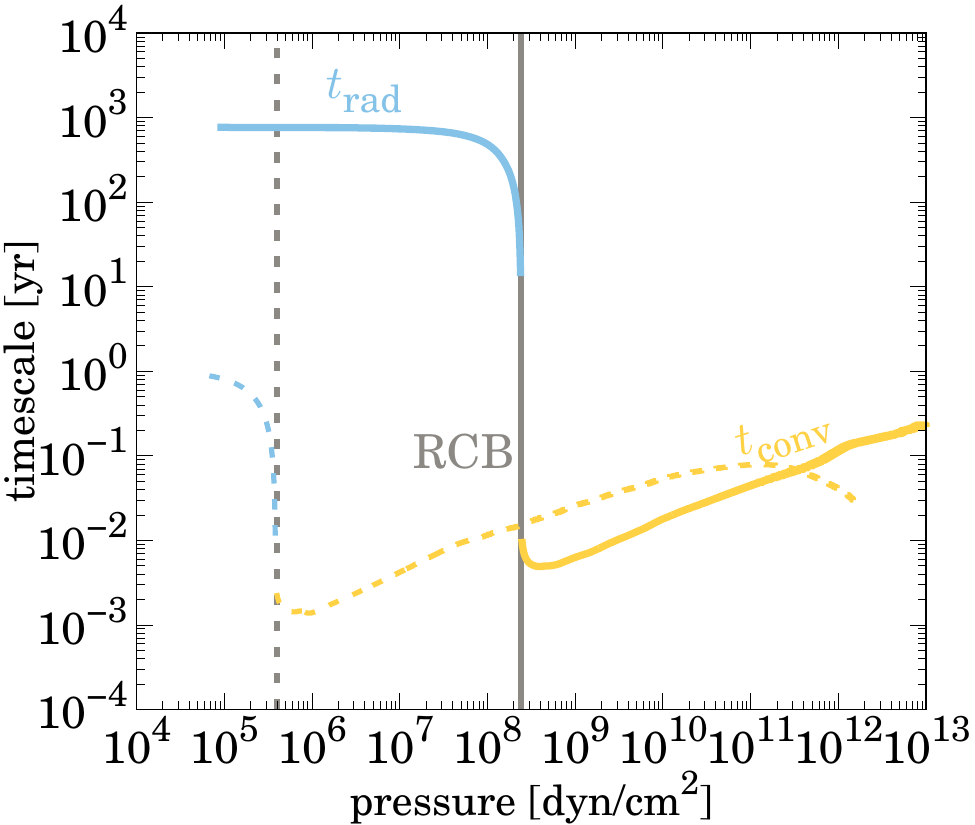}
\caption{Timescales across a fiducial hot Jupiter (solid curves; 200 $M_{\oplus}$, 10 $M_{\oplus}$ core, $S{=}8 \ k_{\rm B}/m_{\rm H}$, equilibrium temperature 2040 K, $Z{=}0.02$) and cold sub-Saturn (dashed curves; 11 $M_{\oplus}$, 10 $M_{\oplus}$ core, $S{=}8 \ k_{\rm B}/m_{\rm H}$, equilibrium temperature 288 K, $Z{=}0.02$). The RCB is demarcated with a vertical grey line. Both the convective turnover time $t_{\rm conv}$ and radiative time $t_{\rm rad}$ are $\ll$the cooling time (${\gtrsim}$Gyr) and/or radius puff-up time (${\gtrsim}10^{5}$ yr), so that our stepping through the adiabats approach is justified. \label{figure:timescales}}
\end{figure}

We compare $t_{\rm conv}$ and $t_{\rm rad}$ in Figure \ref{figure:timescales} across the structure of a fiducial hot Jupiter and cold sub-Saturn. The convective and radiative times are indeed $\ll$the cooling times (${\gtrsim}10^{9}$ yr), radius puff-up times found in this paper (${\gtrsim}10^{5}$ yr), and mass loss times found in Paper II (with typical minima ${\gtrsim}\mathrm{a \ few}{\times}10^{4}$ yr) justifying our use of the stepping through the adiabats approach.

\section{Computing Love Numbers}\label{subsection:lovenumber}

We calculate the planetary Love number $k_{2,\rm p}$ following \citep{ste39,batbodlau09}:

\begin{equation}
    k_{2,\rm p}=\frac{3-2\eta}{2+\eta},
\end{equation}

\noindent where $\eta$ follows from integrating the following differential equation from the core surface (since we assume the core is constant density) out to the planet radius:

\begin{equation}\label{equation:love}
    r\frac{d\eta}{dr}+\eta^{2}-\eta-6+6\frac{\rho(r)}{\langle{\rho(r)}\rangle}(\eta+1)=0,
\end{equation}

\noindent where $\langle\rho\rangle$ is the volume-averaged density interior to radial coordinate $r$:

\begin{equation}
    \langle\rho(r)\rangle=\frac{\int_{R_{\rm c}}^{r}4\pi {r^{\prime}}^{2}\rho(r^{\prime})dr^{\prime}}{\frac{4}{3}\pi r^{3}}.
\end{equation}

Equation \ref{equation:love} is integrated using the run of $\rho(r)$ from our planet structure grid.

\section{Entropy of Mixing}

Our calculation of the entropy of mixing must augment the formulae provided by \cite{saucha95} (corrected for typos in entropy derivative and mixing terms) to include contributions from the addition of heavy elements. In this section we outline how we add these corrections.

Following \cite{saucha95} the ideal entropy of mixing is,

\begin{align}\label{equation:Smix_full}
\begin{split}
    \frac{S_{\rm mix}}{k_{\rm B}}&=N\ln N-\Sigma_{i}N_{i}\ln N_{i}\\
    &=(\mathcal{N}_{\rm H}+\mathcal{N}_{\rm He}+\mathcal{N}_{Z})\ln(\mathcal{N}_{\rm H}+\mathcal{N}_{\rm He}+\mathcal{N}_{Z})\\
    &-N_{\rm H_{2}}\ln N_{\rm H_{2}}-N_{\rm H}\ln N_{\rm H}-N_{\rm H^{+}}\ln N_{\rm H^{+}}\\
    &-N_{\rm e}\ln N_{\rm e}-N_{\rm He}\ln N_{\rm He}-N_{\rm He^{+}}\ln N_{\rm He^{+}}\\
    &-N_{\rm He^{2+}}\ln N_{\rm He^{2+}}-N_{Z}\ln N_{Z}\\
\end{split}
\end{align}

\noindent where $\mathcal{N}_{i}$ are total specific number density fractions (e.g. $\mathcal{N}_{\rm H}{=}N_{\rm H_{2}}{+}N_{\rm H}{+}N_{H^{+}}{+}N^{\rm H}_{\rm e}$) and we have added contributions from metals (we ignore possible metal dissociation). Since the individual hydrogen/helium SCvH entropy tables already include entropy of mixing terms for hydrogen/helium in isolation, we only need to include terms for mixing between hydrogen, helium, and metals \cite[see discussion surrounding equation 51 from][]{saucha95}. We are left with,

\begin{align}
\begin{split}
    \frac{S_{\rm mix}}{k_{\rm B}}&=\mathcal{N}_{\rm H}\ln\bigg(1+\frac{\mathcal{N}_{\rm He}}{\mathcal{N}_{\rm H}}+\frac{\mathcal{N}_{Z}}{\mathcal{N}_{\rm H}}\bigg)\\
    &+\mathcal{N}_{\rm He}\ln\bigg(1+\frac{\mathcal{N}_{\rm H}}{\mathcal{N}_{\rm He}}+\frac{\mathcal{N}_{Z}}{\mathcal{N}_{\rm He}}\bigg)\\
    &+\mathcal{N}_{Z}\ln\bigg(1+\frac{\mathcal{N}_{\rm H}}{\mathcal{N}_{Z}}+\frac{\mathcal{N}_{\rm He}}{\mathcal{N}_{Z}}\bigg)\\
    &+\mathcal{N}_{Z}\ln\mathcal{N}_{Z}-N_{\rm e}\ln N_{\rm e}\\
    &+N^{\rm H}_{\rm e}\ln N^{\rm H}_{\rm e}+N^{\rm He}_{\rm e}\ln N^{\rm He}_{\rm e}-N_{\rm Z}\ln N_{\rm Z}.\\
\end{split}
\end{align}

\noindent The $\mathcal{N}_{i}$ terms for hydrogen and helium are evaluated using equations 35 and 36 of SCvH (i.e. $\mathcal{N}_{\rm H}{=}2X/m_{\rm H}(1{+}3x_{\rm H_{2}}{+}x_{\rm H})$ with $X$ the hydrogen mass fraction and $x_{i}$ the concentrations). For metals we take $\mathcal{N}_{Z}{=}Z(1{+}f_{\rm e})/\mu_{Z}m_{\rm H}$, so that the number of metal particles $N_{Z}{=}Z/\mu_{Z}m_{\rm H}$ and electrons $N^{Z}_{\rm e}{=}Z f_{\rm e}/\mu_{Z}m_{\rm H}$, where $f_{\rm e}$ is the charge state tabulated by \texttt{FEOS}.

\acknowledgments

We thank the referee for a constructive and thorough report that significantly improved the paper. T.H. also thanks Andrew Cumming, Doug Lin, and Christoph Mordasini for insightful discussions. The authors acknowledge the MIT Office of Research Computing and Data for providing high performance computing resources that have contributed to the research results reported in this paper. Figures were made with \texttt{gnuplot}. Numerical calculations were performed using scipy \citep[][]{virgomoli20}.

\bibliography{hallatt}{}

\begin{thebibliography}{}
\expandafter\ifx\csname natexlab\endcsname\relax\def\natexlab#1{#1}\fi
\providecommand{\url}[1]{\href{#1}{#1}}
\providecommand{\dodoi}[1]{doi:~\href{http://doi.org/#1}{\nolinkurl{#1}}}
\providecommand{\doeprint}[1]{\href{http://ascl.net/#1}{\nolinkurl{http://ascl.net/#1}}}
\providecommand{\doarXiv}[1]{\href{https://arxiv.org/abs/#1}{\nolinkurl{https://arxiv.org/abs/#1}}}

\bibitem[{{Ahrer} {et~al.}(2025){Ahrer}, {Gandhi}, {Alderson}, {Kirk}, {Teske}, {Booth}, {McDonald}, {Christie}, {Claringbold}, {Nealon}, {Panwar}, {Veras}, {Wakeford}, {Wheatley}, \& {Zamyatina}}]{ahrganald25}
{Ahrer}, E.-M., {Gandhi}, S., {Alderson}, L., {et~al.} 2025, \mnras, \dodoi{10.1093/mnras/staf819}

\bibitem[{{Arras} \& {Bildsten}(2006)}]{arrbil06}
{Arras}, P., \& {Bildsten}, L. 2006, \apj, 650, 394, \dodoi{10.1086/506011}

\bibitem[{{Attia} {et~al.}(2021){Attia}, {Bourrier}, {Eggenberger}, {Mordasini}, {Beust}, \& {Ehrenreich}}]{attbouegg21}
{Attia}, M., {Bourrier}, V., {Eggenberger}, P., {et~al.} 2021, \aap, 647, A40, \dodoi{10.1051/0004-6361/202039452}

\bibitem[{{Banfield} \& {Murray}(1992)}]{banmur92}
{Banfield}, D., \& {Murray}, N. 1992, \icarus, 99, 390, \dodoi{10.1016/0019-1035(92)90155-Z}

\bibitem[{{Batygin} {et~al.}(2009){Batygin}, {Bodenheimer}, \& {Laughlin}}]{batbodlau09}
{Batygin}, K., {Bodenheimer}, P., \& {Laughlin}, G. 2009, \apjl, 704, L49, \dodoi{10.1088/0004-637X/704/1/L49}

\bibitem[{{Bonomo} {et~al.}(2017){Bonomo}, {Desidera}, {Benatti}, {Borsa}, {Crespi}, {Damasso}, {Lanza}, {Sozzetti}, {Lodato}, {Marzari}, {Boccato}, {Claudi}, {Cosentino}, {Covino}, {Gratton}, {Maggio}, {Micela}, {Molinari}, {Pagano}, {Piotto}, {Poretti}, {Smareglia}, {Affer}, {Biazzo}, {Bignamini}, {Esposito}, {Giacobbe}, {H{\'e}brard}, {Malavolta}, {Maldonado}, {Mancini}, {Martinez Fiorenzano}, {Masiero}, {Nascimbeni}, {Pedani}, {Rainer}, \& {Scandariato}}]{bondesben17}
{Bonomo}, A.~S., {Desidera}, S., {Benatti}, S., {et~al.} 2017, \aap, 602, A107, \dodoi{10.1051/0004-6361/201629882}

\bibitem[{{Bourrier} {et~al.}(2018){Bourrier}, {Lovis}, {Beust}, {Ehrenreich}, {Henry}, {Astudillo-Defru}, {Allart}, {Bonfils}, {S{\'e}gransan}, {Delfosse}, {Cegla}, {Wyttenbach}, {Heng}, {Lavie}, \& {Pepe}}]{boulovbeu18}
{Bourrier}, V., {Lovis}, C., {Beust}, H., {et~al.} 2018, \nat, 553, 477, \dodoi{10.1038/nature24677}

\bibitem[{{Bourrier} {et~al.}(2023){Bourrier}, {Attia}, {Mallonn}, {Marret}, {Lendl}, {Konig}, {Krenn}, {Cretignier}, {Allart}, {Henry}, {Bryant}, {Leleu}, {Nielsen}, {Hebrard}, {Hara}, {Ehrenreich}, {Seidel}, {dos Santos}, {Lovis}, {Bayliss}, {Cegla}, {Dumusque}, {Boisse}, {Boucher}, {Bouchy}, {Pepe}, {Lavie}, {Rey Cerda}, {S{\'e}gransan}, {Udry}, \& {Vrignaud}}]{bouattmall23}
{Bourrier}, V., {Attia}, M., {Mallonn}, M., {et~al.} 2023, \aap, 669, A63, \dodoi{10.1051/0004-6361/202245004}

\bibitem[{{Cassisi} {et~al.}(2007){Cassisi}, {Potekhin}, {Pietrinferni}, {Catelan}, \& {Salaris}}]{caspotpie07}
{Cassisi}, S., {Potekhin}, A.~Y., {Pietrinferni}, A., {Catelan}, M., \& {Salaris}, M. 2007, \apj, 661, 1094, \dodoi{10.1086/516819}

\bibitem[{{Castro-Gonz{\'a}lez} {et~al.}(2024{\natexlab{a}}){Castro-Gonz{\'a}lez}, {Bourrier}, {Lillo-Box}, {Delisle}, {Armstrong}, {Barrado}, \& {Correia}}]{casboulil24}
{Castro-Gonz{\'a}lez}, A., {Bourrier}, V., {Lillo-Box}, J., {et~al.} 2024{\natexlab{a}}, \aap, 689, A250, \dodoi{10.1051/0004-6361/202450957}

\bibitem[{{Castro-Gonz{\'a}lez} {et~al.}(2024{\natexlab{b}}){Castro-Gonz{\'a}lez}, {Lillo-Box}, {Armstrong}, {Acu{\~n}a}, {Aguichine}, {Bourrier}, {Gandhi}, {Sousa}, {Delgado-Mena}, {Moya}, {Adibekyan}, {Correia}, {Barrado}, {Damasso}, {Winn}, {Santos}, {Barkaoui}, {Barros}, {Benkhaldoun}, {Bouchy}, {Brice{\~n}o}, {Caldwell}, {Collins}, {Essack}, {Ghachoui}, {Gillon}, {Hounsell}, {Jehin}, {Jenkins}, {Keniger}, {Law}, {Mann}, {Nielsen}, {Pozuelos}, {Schanche}, {Seager}, {Tan}, {Timmermans}, {Villase{\~n}or}, {Watkins}, \& {Ziegler}}]{caslilarm24}
{Castro-Gonz{\'a}lez}, A., {Lillo-Box}, J., {Armstrong}, D.~J., {et~al.} 2024{\natexlab{b}}, \aap, 691, A233, \dodoi{10.1051/0004-6361/202451656}

\bibitem[{{Chabrier} {et~al.}(2019){Chabrier}, {Mazevet}, \& {Soubiran}}]{chamazsou19}
{Chabrier}, G., {Mazevet}, S., \& {Soubiran}, F. 2019, \apj, 872, 51, \dodoi{10.3847/1538-4357/aaf99f}

\bibitem[{{Chandrasekhar}(1939)}]{cha39}
{Chandrasekhar}, S. 1939, {An introduction to the study of stellar structure} (The University of Chicago press)

\bibitem[{{Choi} {et~al.}(2016){Choi}, {Dotter}, {Conroy}, {Cantiello}, {Paxton}, \& {Johnson}}]{chodotcon16}
{Choi}, J., {Dotter}, A., {Conroy}, C., {et~al.} 2016, \apj, 823, 102, \dodoi{10.3847/0004-637X/823/2/102}

\bibitem[{{Correia} {et~al.}(2020){Correia}, {Bourrier}, \& {Delisle}}]{corboudes20}
{Correia}, A.~C.~M., {Bourrier}, V., \& {Delisle}, J.~B. 2020, \aap, 635, A37, \dodoi{10.1051/0004-6361/201936967}

\bibitem[{{Cox} \& {Giuli}(1968)}]{coxgiu68}
{Cox}, J.~P., \& {Giuli}, R.~T. 1968, {Principles of Stellar Structure} (Cambridge Scientific Publishers ltd)

\bibitem[{{Dermott}(1979)}]{der79}
{Dermott}, S.~F. 1979, \icarus, 37, 310, \dodoi{10.1016/0019-1035(79)90137-4}

\bibitem[{{Dorn} \& {Lichtenberg}(2021)}]{dorlic21}
{Dorn}, C., \& {Lichtenberg}, T. 2021, \apjl, 922, L4, \dodoi{10.3847/2041-8213/ac33af}

\bibitem[{{Doyle} {et~al.}(2025){Doyle}, {Armstrong}, {Acu{\~n}a}, {Osborn}, {Sousa}, {Castro-Gonz{\'a}lez}, {Bourrier}, {Alves}, {Barrado}, {Barros}, {Bayliss}, {Cui}, {Demangeon}, {D{\'\i}az}, {Dumusque}, {Eeles-Nolle}, {Gill}, {Hacker}, {Jenkins}, {Keniger}, {Lafarga}, {Lillo-Box}, {Lockley}, {Nielsen}, {Parc}, {Rodrigues}, {Santerne}, {Santos}, \& {Wheatley}}]{doyarmacu25}
{Doyle}, L., {Armstrong}, D.~J., {Acu{\~n}a}, L., {et~al.} 2025, \mnras, \dodoi{10.1093/mnras/staf670}

\bibitem[{{Dreizler} {et~al.}(2020){Dreizler}, {Crossfield}, {Kossakowski}, {Plavchan}, {Jeffers}, {Kemmer}, {Luque}, {Espinoza}, {Pall{\'e}}, {Stassun}, {Matthews}, {Cale}, {Caballero}, {Schlecker}, {Lillo-Box}, {Zechmeister}, {Lalitha}, {Reiners}, {Soubkiou}, {Bitsch}, {Zapatero Osorio}, {Chaturvedi}, {Hatzes}, {Ricker}, {Vanderspek}, {Latham}, {Seager}, {Winn}, {Jenkins}, {Aceituno}, {Amado}, {Barkaoui}, {Barbieri}, {Batalha}, {Bauer}, {Benneke}, {Benkhaldoun}, {Beichman}, {Berberian}, {Burt}, {Butler}, {Caldwell}, {Chintada}, {Chontos}, {Christiansen}, {Ciardi}, {Cifuentes}, {Collins}, {Collins}, {Combs}, {Cort{\'e}s-Contreras}, {Crane}, {Daylan}, {Dragomir}, {Esparza-Borges}, {Evans}, {Feng}, {Flowers}, {Fukui}, {Fulton}, {Furlan}, {Gaidos}, {Geneser}, {Giacalone}, {Gillon}, {Gonzales}, {Gorjian}, {Hellier}, {Hidalgo}, {Howard}, {Howell}, {Huber}, {Isaacson}, {Jehin}, {Jensen}, {Kaminski}, {Kane}, {Kawauchi}, {Kielkopf}, {Klahr}, {Kosiarek}, {Kreidberg}, {K{\"u}rster}, {Lafarga}, {Livingston}, {Louie},
  {Mann}, {Madrigal-Aguado}, {Matson}, {Mocnik}, {Morales}, {Muirhead}, {Murgas}, {Nandakumar}, {Narita}, {Nowak}, {Oshagh}, {Parviainen}, {Passegger}, {Pollacco}, {Pozuelos}, {Quirrenbach}, {Reefe}, {Ribas}, {Robertson}, {Rodr{\'\i}guez-L{\'o}pez}, {Rose}, {Roy}, {Schweitzer}, {Schlieder}, {Shectman}, {Tanner}, {{\c{S}}enavc{\i}}, {Teske}, {Twicken}, {Villasenor}, {Wang}, {Weiss}, {Wittrock}, {Y{\i}lmaz}, \& {Zohrabi}}]{drecrokos20}
{Dreizler}, S., {Crossfield}, I.~J.~M., {Kossakowski}, D., {et~al.} 2020, \aap, 644, A127, \dodoi{10.1051/0004-6361/202038016}

\bibitem[{{Espinoza-Retamal} {et~al.}(2024){Espinoza-Retamal}, {Stef{\'a}nsson}, {Petrovich}, {Brahm}, {Jord{\'a}n}, {Sedaghati}, {Lucero}, {Tala Pinto}, {Mu{\~n}oz}, {Boyle}, {Leiva}, \& {Suc}}]{espstegud24}
{Espinoza-Retamal}, J.~I., {Stef{\'a}nsson}, G., {Petrovich}, C., {et~al.} 2024, \aj, 168, 185, \dodoi{10.3847/1538-3881/ad70b8}

\bibitem[{{Fabrycky} \& {Tremaine}(2007)}]{fabtre07}
{Fabrycky}, D., \& {Tremaine}, S. 2007, \apj, 669, 1298, \dodoi{10.1086/521702}

\bibitem[{{Faik} {et~al.}(2018){Faik}, {Tauschwitz}, \& {Iosilevskiy}}]{faitauios18}
{Faik}, S., {Tauschwitz}, A., \& {Iosilevskiy}, I. 2018, Computer Physics Communications, 227, 117, \dodoi{10.1016/j.cpc.2018.01.008}

\bibitem[{{Ferguson} {et~al.}(2005){Ferguson}, {Alexander}, {Allard}, {Barman}, {Bodnarik}, {Hauschildt}, {Heffner-Wong}, \& {Tamanai}}]{feraleall05}
{Ferguson}, J.~W., {Alexander}, D.~R., {Allard}, F., {et~al.} 2005, \apj, 623, 585, \dodoi{10.1086/428642}

\bibitem[{{Fortney} \& {Hubbard}(2003)}]{forhub03}
{Fortney}, J.~J., \& {Hubbard}, W.~B. 2003, \icarus, 164, 228, \dodoi{10.1016/S0019-1035(03)00130-1}

\bibitem[{{Fortney} {et~al.}(2007){Fortney}, {Marley}, \& {Barnes}}]{formarbar07}
{Fortney}, J.~J., {Marley}, M.~S., \& {Barnes}, J.~W. 2007, \apj, 659, 1661, \dodoi{10.1086/512120}

\bibitem[{{Freedman} {et~al.}(2014){Freedman}, {Lustig-Yaeger}, {Fortney}, {Lupu}, {Marley}, \& {Lodders}}]{freelusfor14}
{Freedman}, R.~S., {Lustig-Yaeger}, J., {Fortney}, J.~J., {et~al.} 2014, \apjs, 214, 25, \dodoi{10.1088/0067-0049/214/2/25}

\bibitem[{{Freedman} {et~al.}(2008){Freedman}, {Marley}, \& {Lodders}}]{fremarlod08}
{Freedman}, R.~S., {Marley}, M.~S., \& {Lodders}, K. 2008, \apjs, 174, 504, \dodoi{10.1086/521793}

\bibitem[{{Glanz} {et~al.}(2022){Glanz}, {Rozner}, {Perets}, \& {Grishin}}]{glarozper22}
{Glanz}, H., {Rozner}, M., {Perets}, H.~B., \& {Grishin}, E. 2022, \apj, 931, 11, \dodoi{10.3847/1538-4357/ac6807}

\bibitem[{{Goldreich} \& {Nicholson}(1989)}]{golnic89}
{Goldreich}, P., \& {Nicholson}, P.~D. 1989, \apj, 342, 1079, \dodoi{10.1086/167665}

\bibitem[{{Goodman} \& {Oh}(1997)}]{goooh97}
{Goodman}, J., \& {Oh}, S.~P. 1997, \apj, 486, 403, \dodoi{10.1086/304505}

\bibitem[{{Gordon} {et~al.}(1994){Gordon}, {McBride}, \& {Zeleznik}}]{gormcb94}
{Gordon}, S., {McBride}, B., \& {Zeleznik}, F.~J. 1994, {Computer program for calculation of complex chemical equilibrium compositions and applications.} (NASA)

\bibitem[{{Greene} {et~al.}(2016){Greene}, {Line}, {Montero}, {Fortney}, {Lustig-Yaeger}, \& {Luther}}]{grelinmon16}
{Greene}, T.~P., {Line}, M.~R., {Montero}, C., {et~al.} 2016, \apj, 817, 17, \dodoi{10.3847/0004-637X/817/1/17}

\bibitem[{{Guillot}(2010)}]{gui10}
{Guillot}, T. 2010, \aap, 520, A27, \dodoi{10.1051/0004-6361/200913396}

\bibitem[{{Gupta} {et~al.}(2025){Gupta}, {Stixrude}, \& {Schlichting}}]{gupstisch25}
{Gupta}, A., {Stixrude}, L., \& {Schlichting}, H.~E. 2025, \apjl, 982, L35, \dodoi{10.3847/2041-8213/adb631}

\bibitem[{{Haldemann} {et~al.}(2024){Haldemann}, {Dorn}, {Venturini}, {Alibert}, \& {Benz}}]{haldorven24}
{Haldemann}, J., {Dorn}, C., {Venturini}, J., {Alibert}, Y., \& {Benz}, W. 2024, \aap, 681, A96, \dodoi{10.1051/0004-6361/202346965}

\bibitem[{{Hallatt} \& {Lee}(2022)}]{hallee22}
{Hallatt}, T., \& {Lee}, E.~J. 2022, \apj, 924, 9, \dodoi{10.3847/1538-4357/ac32c9}

\bibitem[{{Hansen} \& {Kawaler}(1994)}]{hankaw94}
{Hansen}, C.~J., \& {Kawaler}, S.~D. 1994, {Stellar Interiors. Physical Principles, Structure, and Evolution.} (Springer), \dodoi{10.1007/978-1-4419-9110-2}

\bibitem[{{Helled} \& {Stevenson}(2024)}]{helste24}
{Helled}, R., \& {Stevenson}, D.~J. 2024, AGU Advances, 5, e2024AV001171, \dodoi{10.1029/2024AV001171}

\bibitem[{{Hellier} {et~al.}(2019){Hellier}, {Anderson}, {Triaud}, {Bouchy}, {Burdanov}, {Collier Cameron}, {Delrez}, {Ehrenreich}, {Gillon}, {Jehin}, {Lendl}, {Linder}, {Nielsen}, {Maxted}, {Pepe}, {Pollacco}, {Queloz}, {S{\'e}gransan}, {Smalley}, {Spake}, {Temple}, {Udry}, {West}, \& {Wyttenbach}}]{helandtri19}
{Hellier}, C., {Anderson}, D.~R., {Triaud}, A.~H.~M.~J., {et~al.} 2019, \mnras, 488, 3067, \dodoi{10.1093/mnras/stz1903}

\bibitem[{{Howard} {et~al.}(2025){Howard}, {Helled}, \& {M{\"u}ller}}]{howhelmul25}
{Howard}, S., {Helled}, R., \& {M{\"u}ller}, S. 2025, \aap, 693, L7, \dodoi{10.1051/0004-6361/202452783}

\bibitem[{{Huang} \& {Cumming}(2012)}]{huacum12}
{Huang}, X., \& {Cumming}, A. 2012, \apj, 757, 47, \dodoi{10.1088/0004-637X/757/1/47}

\bibitem[{{Hubbard}(1977)}]{hub77}
{Hubbard}, W.~B. 1977, \icarus, 30, 305, \dodoi{10.1016/0019-1035(77)90164-6}

\bibitem[{{Hubeny} {et~al.}(2003){Hubeny}, {Burrows}, \& {Sudarsky}}]{hubbursud03}
{Hubeny}, I., {Burrows}, A., \& {Sudarsky}, D. 2003, \apj, 594, 1011, \dodoi{10.1086/377080}

\bibitem[{{Hubeny} \& {Mihalas}(2014)}]{hubmih14}
{Hubeny}, I., \& {Mihalas}, D. 2014, {Theory of Stellar Atmospheres} (Princeton University Press)

\bibitem[{{Hut}(1981)}]{hut81}
{Hut}, P. 1981, \aap, 99, 126

\bibitem[{{Ibgui} \& {Burrows}(2009)}]{ibgbur09}
{Ibgui}, L., \& {Burrows}, A. 2009, \apj, 700, 1921, \dodoi{10.1088/0004-637X/700/2/1921}

\bibitem[{{Ito} \& {Ohtsuka}(2019)}]{itooht19}
{Ito}, T., \& {Ohtsuka}, K. 2019, Monographs on Environment, Earth and Planets, 7, 1, \dodoi{10.5047/meep.2019.00701.0001}

\bibitem[{{Jenkins} {et~al.}(2025){Jenkins}, {Vanderburg}, {Sethi}, {Millholland}, {Rodriguez}, {Fossati}, {Krenn}, {Pass}, {Venner}, {Butler}, {Osborn}, {Householder}, {Ziegler}, {Becker}, {Berlind}, {Broeg}, {Calkins}, {Crane}, {Daylan}, {de Wit}, {Eastman}, {Ehrenreich}, {Esquerdo}, {Fausnaugh}, {Gaibor}, {Latham}, {Lendl}, {Mayo}, {Scandariato}, {Shectman}, {Striegel}, {Teske}, \& {Wilson}}]{jensydset25}
{Jenkins}, S., {Vanderburg}, A., {Sethi}, R., {et~al.} 2025, arXiv e-prints, arXiv:2505.10324, \dodoi{10.48550/arXiv.2505.10324}

\bibitem[{{Jin} {et~al.}(2014){Jin}, {Mordasini}, {Parmentier}, {van Boekel}, {Henning}, \& {Ji}}]{jinmorpar14}
{Jin}, S., {Mordasini}, C., {Parmentier}, V., {et~al.} 2014, \apj, 795, 65, \dodoi{10.1088/0004-637X/795/1/65}

\bibitem[{{Jord{\'a}n} {et~al.}(2020){Jord{\'a}n}, {Bakos}, {Bayliss}, {Bento}, {Bhatti}, {Brahm}, {Csubry}, {Espinoza}, {Hartman}, {Henning}, {Mancini}, {Penev}, {Rabus}, {Sarkis}, {Suc}, {de Val-Borro}, {Zhou}, {Butler}, {Teske}, {Crane}, {Shectman}, {Tan}, {Thompson}, {Wallace}, {L{\'a}z{\'a}r}, {Papp}, \& {S{\'a}ri}}]{jorbakbay20}
{Jord{\'a}n}, A., {Bakos}, G.~{\'A}., {Bayliss}, D., {et~al.} 2020, \aj, 160, 222, \dodoi{10.3847/1538-3881/aba530}

\bibitem[{{Kippenhahn} \& {Weigert}(1990)}]{kipwei90}
{Kippenhahn}, R., \& {Weigert}, A. 1990, {Stellar Structure and Evolution} ({Springer})

\bibitem[{{Kirk} {et~al.}(2024){Kirk}, {Ahrer}, {Penzlin}, {Owen}, {Booth}, {Alderson}, {Christie}, {Claringbold}, {Esparza-Borges}, {Fisher}, {L{\'o}pez-Morales}, {Mayne}, {McCormack}, {Meech}, {Panwar}, {Powell}, {Sergeev}, {Taylor}, {Tsai}, {Valentine}, {Wakeford}, {Wheatley}, \& {Zamyatina}}]{kirahrpen24}
{Kirk}, J., {Ahrer}, E.-M., {Penzlin}, A. B.~T., {et~al.} 2024, RAS Techniques and Instruments, 3, 691, \dodoi{10.1093/rasti/rzae043}

\bibitem[{{Knierim} \& {Helled}(2024)}]{knihel24}
{Knierim}, H., \& {Helled}, R. 2024, \apj, 977, 227, \dodoi{10.3847/1538-4357/ad8dd0}

\bibitem[{{Komacek} \& {Youdin}(2017)}]{komyou17}
{Komacek}, T.~D., \& {Youdin}, A.~N. 2017, \apj, 844, 94, \dodoi{10.3847/1538-4357/aa7b75}

\bibitem[{{Koskinen} {et~al.}(2022){Koskinen}, {Lavvas}, {Huang}, {Bergsten}, {Fernandes}, \& {Young}}]{koslavhua22}
{Koskinen}, T.~T., {Lavvas}, P., {Huang}, C., {et~al.} 2022, \apj, 929, 52, \dodoi{10.3847/1538-4357/ac4f45}

\bibitem[{{Lai}(2012)}]{lai12}
{Lai}, D. 2012, \mnras, 423, 486, \dodoi{10.1111/j.1365-2966.2012.20893.x}

\bibitem[{{Leconte} {et~al.}(2010){Leconte}, {Chabrier}, {Baraffe}, \& {Levrard}}]{lecchabar10}
{Leconte}, J., {Chabrier}, G., {Baraffe}, I., \& {Levrard}, B. 2010, \aap, 516, A64, \dodoi{10.1051/0004-6361/201014337}

\bibitem[{{Lee} {et~al.}(2014){Lee}, {Chiang}, \& {Ormel}}]{leechiorm14}
{Lee}, E.~J., {Chiang}, E., \& {Ormel}, C.~W. 2014, \apj, 797, 95, \dodoi{10.1088/0004-637X/797/2/95}

\bibitem[{{Lopez} \& {Fortney}(2014)}]{lopfor14}
{Lopez}, E.~D., \& {Fortney}, J.~J. 2014, \apj, 792, 1, \dodoi{10.1088/0004-637X/792/1/1}

\bibitem[{{Lopez} {et~al.}(2012){Lopez}, {Fortney}, \& {Miller}}]{lopformil12}
{Lopez}, E.~D., {Fortney}, J.~J., \& {Miller}, N. 2012, \apj, 761, 59, \dodoi{10.1088/0004-637X/761/1/59}

\bibitem[{{Lu} {et~al.}(2025){Lu}, {An}, {Li}, {Millholland}, {Rice}, {Brandt}, \& {Brandt}}]{luanli25}
{Lu}, T., {An}, Q., {Li}, G., {et~al.} 2025, \apj, 979, 218, \dodoi{10.3847/1538-4357/ad9b79}

\bibitem[{{Lubow} {et~al.}(1997){Lubow}, {Tout}, \& {Livio}}]{lubtouliv97}
{Lubow}, S.~H., {Tout}, C.~A., \& {Livio}, M. 1997, \apj, 484, 866, \dodoi{10.1086/304369}

\bibitem[{{Mankovich} \& {Fuller}(2021)}]{manful21}
{Mankovich}, C.~R., \& {Fuller}, J. 2021, Nature Astronomy, 5, 1103, \dodoi{10.1038/s41550-021-01448-3}

\bibitem[{{Markham} {et~al.}(2022){Markham}, {Guillot}, \& {Stevenson}}]{marguiste22}
{Markham}, S., {Guillot}, T., \& {Stevenson}, D. 2022, \aap, 665, A12, \dodoi{10.1051/0004-6361/202243359}

\bibitem[{{Marleau} \& {Cumming}(2014)}]{marcum14}
{Marleau}, G.~D., \& {Cumming}, A. 2014, \mnras, 437, 1378, \dodoi{10.1093/mnras/stt1967}

\bibitem[{{Matsumura} {et~al.}(2010){Matsumura}, {Peale}, \& {Rasio}}]{matpearas10}
{Matsumura}, S., {Peale}, S.~J., \& {Rasio}, F.~A. 2010, \apj, 725, 1995, \dodoi{10.1088/0004-637X/725/2/1995}

\bibitem[{{Miller} {et~al.}(2009){Miller}, {Fortney}, \& {Jackson}}]{milforjac09}
{Miller}, N., {Fortney}, J.~J., \& {Jackson}, B. 2009, \apj, 702, 1413, \dodoi{10.1088/0004-637X/702/2/1413}

\bibitem[{{Millholland}(2019)}]{mil19}
{Millholland}, S. 2019, \apj, 886, 72, \dodoi{10.3847/1538-4357/ab4c3f}

\bibitem[{{Millholland} {et~al.}(2020){Millholland}, {Petigura}, \& {Batygin}}]{milpetbat20}
{Millholland}, S., {Petigura}, E., \& {Batygin}, K. 2020, \apj, 897, 7, \dodoi{10.3847/1538-4357/ab959c}

\bibitem[{{Movshovitz} \& {Podolak}(2008)}]{movpod08}
{Movshovitz}, N., \& {Podolak}, M. 2008, \icarus, 194, 368, \dodoi{10.1016/j.icarus.2007.09.018}

\bibitem[{{M{\"u}ller} {et~al.}(2020{\natexlab{a}}){M{\"u}ller}, {Ben-Yami}, \& {Helled}}]{mulbenhel20}
{M{\"u}ller}, S., {Ben-Yami}, M., \& {Helled}, R. 2020{\natexlab{a}}, \apj, 903, 147, \dodoi{10.3847/1538-4357/abba19}

\bibitem[{{M{\"u}ller} \& {Helled}(2021)}]{mulhel21}
{M{\"u}ller}, S., \& {Helled}, R. 2021, \mnras, 507, 2094, \dodoi{10.1093/mnras/stab2250}

\bibitem[{{M{\"u}ller} {et~al.}(2020{\natexlab{b}}){M{\"u}ller}, {Helled}, \& {Cumming}}]{mulhelcum20}
{M{\"u}ller}, S., {Helled}, R., \& {Cumming}, A. 2020{\natexlab{b}}, \aap, 638, A121, \dodoi{10.1051/0004-6361/201937376}

\bibitem[{{Murphy} {et~al.}(2025){Murphy}, {Beatty}, {Welbanks}, \& {Fu}}]{murbeawel25}
{Murphy}, M.~M., {Beatty}, T.~G., {Welbanks}, L., \& {Fu}, G. 2025, \aj, 169, 286, \dodoi{10.3847/1538-3881/adc684}

\bibitem[{{Naoz}(2016)}]{nao16}
{Naoz}, S. 2016, \araa, 54, 441, \dodoi{10.1146/annurev-astro-081915-023315}

\bibitem[{{Naponiello} {et~al.}(2025){Naponiello}, {Vissapragada}, {Bonomo}, {Steinmeyer}, {Filomeno}, {D'Orazi}, {Dorn}, {Sozzetti}, {Mancini}, {Lanza}, {Biazzo}, {Watkins}, {H{\'e}brard}, {Lissauer}, {Howell}, {Ciardi}, {Mantovan}, {Baker}, {Bourrier}, {Buchhave}, {Clark}, {Collins}, {Cosentino}, {Damasso}, {Dumusque}, {Fiorenzano}, {Forveille}, {Heidari}, {Latham}, {Littlefield}, {L{\'o}pez-Morales}, {Lund}, {Malavolta}, {Manni}, {Nardiello}, {Pinamonti}, {Yee}, {Zambelli}, {Ziegler}, \& {Zingales}}]{napvisbon25}
{Naponiello}, L., {Vissapragada}, S., {Bonomo}, A.~S., {et~al.} 2025, arXiv e-prints, arXiv:2505.10123, \dodoi{10.48550/arXiv.2505.10123}

\bibitem[{{{\"O}berg} {et~al.}(2011){{\"O}berg}, {Murray-Clay}, \& {Bergin}}]{obemurber11}
{{\"O}berg}, K.~I., {Murray-Clay}, R., \& {Bergin}, E.~A. 2011, \apjl, 743, L16, \dodoi{10.1088/2041-8205/743/1/L16}

\bibitem[{{Ogilvie}(2014)}]{ogi14}
{Ogilvie}, G.~I. 2014, \araa, 52, 171, \dodoi{10.1146/annurev-astro-081913-035941}

\bibitem[{{Ogilvie} \& {Lin}(2004)}]{ogilin04}
{Ogilvie}, G.~I., \& {Lin}, D.~N.~C. 2004, \apj, 610, 477, \dodoi{10.1086/421454}

\bibitem[{{Owen} \& {Wu}(2016)}]{owewu16}
{Owen}, J.~E., \& {Wu}, Y. 2016, \apj, 817, 107, \dodoi{10.3847/0004-637X/817/2/107}

\bibitem[{{Paczy{\'n}ski} \& {Sienkiewicz}(1972)}]{pacsie72}
{Paczy{\'n}ski}, B., \& {Sienkiewicz}, R. 1972, \actaa, 22, 73

\bibitem[{{Paxton} {et~al.}(2011){Paxton}, {Bildsten}, {Dotter}, {Herwig}, {Lesaffre}, \& {Timmes}}]{paxbildot11}
{Paxton}, B., {Bildsten}, L., {Dotter}, A., {et~al.} 2011, \apjs, 192, 3, \dodoi{10.1088/0067-0049/192/1/3}

\bibitem[{{Paxton} {et~al.}(2013){Paxton}, {Cantiello}, {Arras}, {Bildsten}, {Brown}, {Dotter}, {Mankovich}, {Montgomery}, {Stello}, {Timmes}, \& {Townsend}}]{paxcanarr13}
{Paxton}, B., {Cantiello}, M., {Arras}, P., {et~al.} 2013, \apjs, 208, 4, \dodoi{10.1088/0067-0049/208/1/4}

\bibitem[{{Paxton} {et~al.}(2015){Paxton}, {Marchant}, {Schwab}, {Bauer}, {Bildsten}, {Cantiello}, {Dessart}, {Farmer}, {Hu}, {Langer}, {Townsend}, {Townsley}, \& {Timmes}}]{paxmarsch15}
{Paxton}, B., {Marchant}, P., {Schwab}, J., {et~al.} 2015, \apjs, 220, 15, \dodoi{10.1088/0067-0049/220/1/15}

\bibitem[{{Paxton} {et~al.}(2018){Paxton}, {Schwab}, {Bauer}, {Bildsten}, {Blinnikov}, {Duffell}, {Farmer}, {Goldberg}, {Marchant}, {Sorokina}, {Thoul}, {Townsend}, \& {Timmes}}]{paxschbau18}
{Paxton}, B., {Schwab}, J., {Bauer}, E.~B., {et~al.} 2018, \apjs, 234, 34, \dodoi{10.3847/1538-4365/aaa5a8}

\bibitem[{{Paxton} {et~al.}(2019){Paxton}, {Smolec}, {Schwab}, {Gautschy}, {Bildsten}, {Cantiello}, {Dotter}, {Farmer}, {Goldberg}, {Jermyn}, {Kanbur}, {Marchant}, {Thoul}, {Townsend}, {Wolf}, {Zhang}, \& {Timmes}}]{paxsmosch19}
{Paxton}, B., {Smolec}, R., {Schwab}, J., {et~al.} 2019, \apjs, 243, 10, \dodoi{10.3847/1538-4365/ab2241}

\bibitem[{{Penzlin} {et~al.}(2024){Penzlin}, {Booth}, {Kirk}, {Owen}, {Ahrer}, {Christie}, {Claringbold}, {Esparza-Borges}, {L{\'o}pez-Morales}, {Mayne}, {McCormack}, {Meech}, {Panwar}, {Powell}, {Sergeev}, {Taylor}, {Wheatley}, \& {Zamyatina}}]{penbookir24}
{Penzlin}, A. B.~T., {Booth}, R.~A., {Kirk}, J., {et~al.} 2024, \mnras, 535, 171, \dodoi{10.1093/mnras/stae2362}

\bibitem[{{Petigura} {et~al.}(2018){Petigura}, {Marcy}, {Winn}, {Weiss}, {Fulton}, {Howard}, {Sinukoff}, {Isaacson}, {Morton}, \& {Johnson}}]{petmarwin18}
{Petigura}, E.~A., {Marcy}, G.~W., {Winn}, J.~N., {et~al.} 2018, \aj, 155, 89, \dodoi{10.3847/1538-3881/aaa54c}

\bibitem[{{Petrovich}(2015)}]{pet15a}
{Petrovich}, C. 2015, \apj, 799, 27, \dodoi{10.1088/0004-637X/799/1/27}

\bibitem[{{Polanski} {et~al.}(2024){Polanski}, {Lubin}, {Beard}, {Akana Murphy}, {Rubenzahl}, {Hill}, {Crossfield}, {Chontos}, {Robertson}, {Isaacson}, {Kane}, {Ciardi}, {Batalha}, {Dressing}, {Fulton}, {Howard}, {Huber}, {Petigura}, {Weiss}, {Angelo}, {Behmard}, {Blunt}, {Brinkman}, {Dai}, {Dalba}, {Fetherolf}, {Giacalone}, {Hirsch}, {Holcomb}, {Kosiarek}, {Mayo}, {MacDougall}, {Mo{\v{c}}nik}, {Pidhorodetska}, {Rice}, {Rosenthal}, {Scarsdale}, {Turtelboom}, {Tyler}, {Van Zandt}, {Yee}, {Coria}, {Dulz}, {Hartman}, {Householder}, {Lange}, {Langford}, {Louden}, {Siegel}, {Gilbert}, {Gonzales}, {Schlieder}, {Boyle}, {Christiansen}, {Clark}, {Fernandes}, {Lund}, {Savel}, {Gill}, {Beichman}, {Matson}, {Matthews}, {Furlan}, {Howell}, {Scott}, {Everett}, {Livingston}, {Ershova}, {Cheryasov}, {Safonov}, {Lillo-Box}, {Barrado}, \& {Morales-Calder{\'o}n}}]{pollubbea24}
{Polanski}, A.~S., {Lubin}, J., {Beard}, C., {et~al.} 2024, \apjs, 272, 32, \dodoi{10.3847/1538-4365/ad4484}

\bibitem[{{Potekhin} {et~al.}(2015){Potekhin}, {Pons}, \& {Page}}]{potponjos15}
{Potekhin}, A.~Y., {Pons}, J.~A., \& {Page}, D. 2015, \ssr, 191, 239, \dodoi{10.1007/s11214-015-0180-9}

\bibitem[{{Ram{\'\i}rez Reyes} {et~al.}(2025){Ram{\'\i}rez Reyes}, {Jenkins}, {Sedaghati}, {Seidel}, {Pavlenko}, {Palle}, {L{\'o}pez-Morales}, {Alves}, {Vines}, {Pe{\~n}a R}, {D{\'\i}az}, \& {Rojo}}]{ramjensed25}
{Ram{\'\i}rez Reyes}, R., {Jenkins}, J.~S., {Sedaghati}, E., {et~al.} 2025, \aap, 695, A26, \dodoi{10.1051/0004-6361/202451044}

\bibitem[{{Rauer} {et~al.}(2025){Rauer}, {Aerts}, {Cabrera}, {Deleuil}, {Erikson}, {Gizon}, {Goupil}, {Heras}, {Walloschek}, {Lorenzo-Alvarez}, {Marliani}, {Martin-Garcia}, {Mas-Hesse}, {O'Rourke}, {Osborn}, {Pagano}, {Piotto}, {Pollacco}, {Ragazzoni}, {Ramsay}, {Udry}, {Appourchaux}, {Benz}, {Brandeker}, {G{\"u}del}, {Janot-Pacheco}, {Kabath}, {Kjeldsen}, {Min}, {Santos}, {Smith}, {Suarez}, {Werner}, {Aboudan}, {Abreu}, {Acu{\~n}a}, {Adams}, {Adibekyan}, {Affer}, {Agneray}, {Agnor}, {Aguirre B{\o}rsen-Koch}, {Ahmed}, {Aigrain}, {Al-Bahlawan}, {Alcacera Gil}, {Alei}, {Alencar}, {Alexander}, {Alfonso-Garz{\'o}n}, {Alibert}, {Allende Prieto}, {Almeida}, {Alonso Sobrino}, {Altavilla}, {Althaus}, {Alvarez Trujillo}, {Amarsi}, {Ammler-von Eiff}, {Am{\^o}res}, {Andrade}, {Antoniadis-Karnavas}, {Ant{\'o}nio}, {Aparicio del Moral}, {Appolloni}, {Arena}, {Armstrong}, {Aroca Aliaga}, {Asplund}, {Audenaert}, {Auricchio}, {Avelino}, {Baeke}, {Bailli{\'e}}, {Balado}, {Ballber Balaguer{\'o}}, {Balestra}, {Ball}, {Ballans},
  {Ballot}, {Barban}, {Barbary}, {Barbieri}, {Barcel{\'o} Forteza}, {Barker}, {Barklem}, {Barnes}, {Barrado Navascues}, {Barragan}, {Baruteau}, {Basu}, {Baudin}, {Baumeister}, {Bayliss}, {Bazot}, {Beck}, {Belkacem}, {Bellinger}, {Benatti}, {Benomar}, {B{\'e}rard}, {Bergemann}, {Bergomi}, {Bernardo}, {Biazzo}, {Bignamini}, {Bigot}, {Billot}, {Binet}, {Biondi}, {Biondi}, {Birch}, {Bitsch}, {Bluhm Ceballos}, {B{\'o}di}, {Bogn{\'a}r}, {Boisse}, {Bolmont}, {Bonanno}, {Bonavita}, {Bonfanti}, {Bonfils}, {Bonito}, {Bonomo}, {B{\"o}rner}, {Boro Saikia}, {Borreguero Mart{\'\i}n}, {Borsa}, {Borsato}, {Bossini}, {Bouchy}, {Bou{\'e}}, {Boufleur}, {Boumier}, {Bourrier}, {Bowman}, {Bozzo}, {Bradley}, {Bray}, {Bressan}, {Breton}, {Brienza}, {Brito}, {Brogi}, {Brown}, {Brown}, {Brun}, {Bruno}, {Bruns}, {Buchhave}, {Bugnet}, {Buldgen}, {Burgess}, {Busatta}, {Busso}, {Buzasi}, {Caballero}, {Cabral}, {Cabrero Gomez}, {Calderone}, {Cameron}, {Cameron}, {Campante}, {Campos Gestal}, {Canto Martins}, {Cara}, {Carone}, {Carrasco},
  {Casagrande}, {Casewell}, {Cassisi}, {Castellani}, {Castro}, {Catala}, {Catal{\'a}n Fern{\'a}ndez}, {Catelan}, {Cegla}, {Cerruti}, {Cessa}, {Chadid}, {Chaplin}, {Charpinet}, {Chiappini}, {Chiarucci}, {Chiavassa}, {Chinellato}, {Chirulli}, {Christensen-Dalsgaard}, {Church}, {Claret}, {Clarke}, {Claudi}, {Clermont}, {Coelho}, {Coelho}, {Cogato}, {Colom{\'e}}, {Condamin}, {Conde Garc{\'\i}a}, \& {Conseil}}]{rauaercab25}
{Rauer}, H., {Aerts}, C., {Cabrera}, J., {et~al.} 2025, Experimental Astronomy, 59, 26, \dodoi{10.1007/s10686-025-09985-9}

\bibitem[{{Rogers}(2025)}]{rog25}
{Rogers}, J.~G. 2025, \mnras, 539, 2230, \dodoi{10.1093/mnras/staf628}

\bibitem[{{Rozner} {et~al.}(2022){Rozner}, {Glanz}, {Perets}, \& {Grishin}}]{rozglaper22}
{Rozner}, M., {Glanz}, H., {Perets}, H.~B., \& {Grishin}, E. 2022, \apj, 931, 10, \dodoi{10.3847/1538-4357/ac6808}

\bibitem[{{Saumon} {et~al.}(1995){Saumon}, {Chabrier}, \& {van Horn}}]{saucha95}
{Saumon}, D., {Chabrier}, G., \& {van Horn}, H.~M. 1995, \apjs, 99, 713, \dodoi{10.1086/192204}

\bibitem[{{Schwarzschild}(1958)}]{sch58}
{Schwarzschild}, M. 1958, {Structure and evolution of the stars.} (Princeton University Press)

\bibitem[{{Sethi} \& {Millholland}(2025)}]{setmil25}
{Sethi}, R., \& {Millholland}, S. 2025, arXiv e-prints, arXiv:2506.24100, \dodoi{10.48550/arXiv.2506.24100}

\bibitem[{{Sterne}(1939)}]{ste39}
{Sterne}, T.~E. 1939, \mnras, 99, 451, \dodoi{10.1093/mnras/99.5.451}

\bibitem[{{Sur} {et~al.}(2024){Sur}, {Su}, {Tejada Arevalo}, {Chen}, \& {Burrows}}]{sursutej24}
{Sur}, A., {Su}, Y., {Tejada Arevalo}, R., {Chen}, Y.-X., \& {Burrows}, A. 2024, \apj, 971, 104, \dodoi{10.3847/1538-4357/ad57c3}

\bibitem[{{Thorngren} {et~al.}(2021){Thorngren}, {Fortney}, {Lopez}, {Berger}, \& {Huber}}]{thoforlop21}
{Thorngren}, D.~P., {Fortney}, J.~J., {Lopez}, E.~D., {Berger}, T.~A., \& {Huber}, D. 2021, \apjl, 909, L16, \dodoi{10.3847/2041-8213/abe86d}

\bibitem[{{Tinetti} {et~al.}(2022){Tinetti}, {Eccleston}, {Lueftinger}, {Salvignol}, {Fahmy}, \& {Alves de Oliveira}}]{tinecclue22}
{Tinetti}, G., {Eccleston}, P., {Lueftinger}, T., {et~al.} 2022, in European Planetary Science Congress, EPSC2022--1114, \dodoi{10.5194/epsc2022-1114}

\bibitem[{{Tittemore} \& {Wisdom}(1990)}]{titwis90}
{Tittemore}, W.~C., \& {Wisdom}, J. 1990, \icarus, 85, 394, \dodoi{10.1016/0019-1035(90)90125-S}

\bibitem[{{Valencia} {et~al.}(2013){Valencia}, {Guillot}, {Parmentier}, \& {Freedman}}]{valguipar13}
{Valencia}, D., {Guillot}, T., {Parmentier}, V., \& {Freedman}, R.~S. 2013, \apj, 775, 10, \dodoi{10.1088/0004-637X/775/1/10}

\bibitem[{{Valluri} {et~al.}(2000){Valluri}, {Jeffrey}, \& {Corless}}]{valjefcor00}
{Valluri}, S.~R., {Jeffrey}, D.~J., \& {Corless}, R.~M. 2000, Canadian Journal of Physics, 78, 823, \dodoi{10.1139/p00-065}

\bibitem[{{Vazan} {et~al.}(2018){Vazan}, {Helled}, \& {Guillot}}]{vazhelgui18}
{Vazan}, A., {Helled}, R., \& {Guillot}, T. 2018, \aap, 610, L14, \dodoi{10.1051/0004-6361/201732522}

\bibitem[{{Vazan} {et~al.}(2013){Vazan}, {Kovetz}, {Podolak}, \& {Helled}}]{vazkovpod13}
{Vazan}, A., {Kovetz}, A., {Podolak}, M., \& {Helled}, R. 2013, \mnras, 434, 3283, \dodoi{10.1093/mnras/stt1248}

\bibitem[{{Virtanen} {et~al.}(2020){Virtanen}, {Gommers}, {Oliphant}, {Haberland}, {Reddy}, {Cournapeau}, {Burovski}, {Peterson}, {Weckesser}, {Bright}, {van der Walt}, {Brett}, {Wilson}, {Millman}, {Mayorov}, {Nelson}, {Jones}, {Kern}, {Larson}, {Carey}, {Polat}, {Feng}, {Moore}, {VanderPlas}, {Laxalde}, {Perktold}, {Cimrman}, {Henriksen}, {Quintero}, {Harris}, {Archibald}, {Ribeiro}, {Pedregosa}, {van Mulbregt}, \& {SciPy 1. 0 Contributors}}]{virgomoli20}
{Virtanen}, P., {Gommers}, R., {Oliphant}, T.~E., {et~al.} 2020, Nature Methods, 17, 261, \dodoi{10.1038/s41592-019-0686-2}

\bibitem[{{Vissapragada} \& {Behmard}(2025)}]{visbeh25}
{Vissapragada}, S., \& {Behmard}, A. 2025, \aj, 169, 117, \dodoi{10.3847/1538-3881/ada143}

\bibitem[{{Wahl} {et~al.}(2017){Wahl}, {Hubbard}, {Militzer}, {Guillot}, {Miguel}, {Movshovitz}, {Kaspi}, {Helled}, {Reese}, {Galanti}, {Levin}, {Connerney}, \& {Bolton}}]{wahhubmil17}
{Wahl}, S.~M., {Hubbard}, W.~B., {Militzer}, B., {et~al.} 2017, \grl, 44, 4649, \dodoi{10.1002/2017GL073160}

\bibitem[{{Xu} \& {Dai}(2025)}]{xudai25}
{Xu}, Y., \& {Dai}, F. 2025, \apj, 981, 142, \dodoi{10.3847/1538-4357/adb281}

\bibitem[{{Yu} \& {Dai}(2024)}]{yudai24}
{Yu}, H., \& {Dai}, F. 2024, \apj, 972, 159, \dodoi{10.3847/1538-4357/ad5ffb}

\end{thebibliography}
\bibliographystyle{aasjournal}

\end{document}